  \documentclass[11pt,letterpaper]{article}
  \usepackage{amstex}
  \usepackage{mathlet}
  \usepackage{epsfig}
  \usepackage{citesort}
\newcounter{Results}

\newcommand{\mbold}[1]{{\mbox{\boldmath{$#1$}\unboldmath}}}
\newcommand{\mcal}[1]{{\mbox{\boldmath{${\mathcal{#1}}$}\unboldmath}}} 
\newcommand{\inner}{\hskip.7mm 
                   \rule[.1mm]{2.3mm}{.16mm}\rule[.1mm]{.16mm}{2.mm}\hskip.7mm}
\newcommand{\D}{{\rm d}}
\newcommand{\DD}{{\rm D}}
\newcommand{\psize}{\hsize}
\newcommand{\acos}{\text{acos}}

\begin{document}

\bibliographystyle{unsrt}

\renewcommand{\thefootnote}{\fnsymbol{footnote}} 

\begin{titlepage}

\rightline{CERN-TH/96-26 (revised)}  
\rightline{Cologne-thp-1996-H06 (revised)} 
\rightline{\mbox{}}
\begin{center}
{\LARGE Volterra Distortions, Spinning Strings, and Cosmic Defects} 
\vspace{10mm}\noindent

{\large
Roland A. Puntigam}$^1$\\ 
Institute for Theoretical Physics, University of Cologne,
D-50923 Cologne, Germany\\ 

\vspace{7mm} 

{\large
Harald H. Soleng}$^{2,3}$\\ 
Theory Division, CERN, CH-1211 Geneva 23, Switzerland

\vspace{5mm} 
{\mbox{ }} 

\end{center}

\vspace{15mm} 

\begin{abstract}%
Cosmic strings, as topological spacetime defects, show striking 
resemblance to defects in solid continua: distortions, which can be
classified into disclinations and dislocations, are line-like defects  
characterized by a delta function-valued curvature and torsion distribution
giving rise to rotational and translational holonomy. We exploit this 
analogy and investigate how distortions can be adapted in a systematic 
manner from solid state systems to Einstein--Cartan gravity. 
As distortions are efficiently described within the framework of 
a $SO(3) {\;{\rlap{$\supset$}\times}\;} T(3)$ gauge theory of solid continua 
with line defects, we are led in a straightforward way to a Poincar\'e gauge 
approach to gravity which is a natural framework for introducing the 
notion of \emph{distorted spacetimes}. Constructing all ten possible 
distorted spacetimes, we recover, inter alia, the well-known exterior 
spacetime of a spin-polarized cosmic string as a special case of such 
a geometry. In a second step, we search for matter distributions which, 
in Einstein--Cartan gravity, act as sources of distorted spacetimes. 
The resulting solutions, appropriately matched to the distorted vacua, 
are cylindrically symmetric and are interpreted as spin-polarized cosmic 
strings and cosmic dislocations.

\vspace{5mm}
\end{abstract}

\vfill

\noindent
CERN-TH/96-26 (revised)\\
February 1996, revised November 1996

\noindent
\footnoterule 

\vspace{0.5cm}

${}^1$
Electronic address: \verb+rap@thp.uni-koeln.de+ 

${}^2$
Electronic address: \verb+harald@nordita.dk+ 

${}^3$
Present address: NORDITA, Blegdamsvej 17, DK-2100 Copenhagen {\O}, Denmark 
\end{titlepage}
 
\renewcommand{\thefootnote}{\arabic{footnote}} 
\addtocounter{footnote}{-\value{footnote}} 
\addtocounter{page}{-\value{page}} 
\addtocounter{page}{2}
 
\section{Introduction}\label{Sect:Introduction}

One of the most interesting features of the spacetime of a straight cosmic 
string \cite{Vilenkin81,Gibbons90,Vilenkin94} in general relativity is 
its conical structure \cite{Hiscock85,Vickers87}. If the string is 
investigated within the Einstein--Cartan theory (EC theory) of 
gravity \cite{Hehl76}, then the geometry of the string spacetime may 
additionally have a chiral structure \cite{Bekenstein92,Letelier95a}. 
More precisely, the string represents a topological defect of spacetime 
that may be described in geometrical terms by delta function-valued 
torsion and curvature components.
  
It has been noticed by several authors that the spacetime geometries of 
cosmic strings in $3+1$ dimensions \cite{Holz92,Gal'tsov93,Edelen94,Tod94} 
exhibit some close relations to so-called {\em distortions\/} of solids, 
which may likewise be regarded as topological defect lines carrying 
torsion and curvature. Similar analogies hold in $2+1$ dimensional
gravity \cite{Holz88,Gerbert90}, where the spacetime geometry of a point 
particle can be understood in terms of distortions. This point has recently
been studied in detail by Kohler \cite{Kohler95,Kohler95b}.

Distortions were introduced by Volterra \cite{Volterra07} in the context
of the theory of elasticity and have later been subject to innumerable 
investigations in the context of both solid continua and crystals,
see, for instance, Nabarro \cite{Nabarro67}, Kl\'eman \cite{Kleman80}, Kr\"oner
\cite{Kroner81}, and references therein. It is remarkable that already Nabarro
mentioned the possible existence of timelike distortions 
\cite[p.~588]{Nabarro67} and the resemblance of Marder's cylindrically 
symmetric solution \cite{Marder59,Marder62} of the Einstein equation 
to distortions.

In this article, we intend to investigate in a  systematic way the above-%
mentioned analogy between distortions of solids and defect structures of 
Riemann--Cartan manifolds. This is done by means of a reexamination of the 
classical Volterra process, see below, which results naturally
in the construction of what we call {\em distortions of spacetime}. 
The straight cosmic string is the most prominent example of such a geometry.

In the EC theory of gravity, we have two kinds of sources: 
energy-momentum, which generates curvature just as in ordinary Einsteinian 
gravity, and spin-angular momentum, generating torsion. If we want to construct
some matter distribution which has as its exterior spacetime 
the defect spacetime under consideration, we can think of to
two inequivalent strategies. Taking the cosmic string as a prototype, 
it may, on the one hand, be considered as a {\em thin string\/} of zero width;
on the other hand, the interior spacetime can be modeled as a 
{\em thick string\/}, i.e. as a cylinder with finite radius. There are 
two reasons for adopting the latter approach in the present investigation.  

First of all, gauge theories that allow for string formation during 
spontaneous symmetry breaking predict extremely thin, but finite strings, 
whose radii depend on the energy scale of symmetry breaking \cite{Vilenkin94}. 
Secondly, it has been shown by Geroch and Traschen \cite{Geroch87} that---due 
to the nonlinearity of the field equations---it is mathematically 
ambiguous to define matter currents corresponding to thin strings 
(see, however, the recent work by Clarke et al. \cite{Clarke96}
for a possibility to deal with two-dimensional distributional sources by 
means of Colombeau's new generalized functions). 
Interior solutions with finite radius have
already been found for massive strings in general relativity 
\cite{Gott85,Linet85} and spin-polarized strings in 
EC theory \cite{Soleng90,Soleng92a}. Here, we use 
this construction to find interior solutions that can be matched to more 
general defect spacetimes.

We have organized this article as follows: in Section~\ref{Sect:Distortions}
we generalize the Volterra process to $3+1$ dimensions using differential 
geometric and gauge theoretic methods and study explicit examples of 
Volterra distorted spacetimes. In detail we discuss only those distortions
that are matched to the corresponding interior matter-filled solution in 
Section \ref{Sect:Extended}. Here we construct interior solutions of the 
Einstein--Cartan field equations, which are matched to the Volterra
defect spacetimes, and the resulting matter distributions are interpreted 
as cosmic strings and cosmic dislocations.

For the sake of notational compactness, we use at the same time Cartesian 
coordinates $\{t, x, y, z\}$ and cylindrical coordinates $\{t, r, \phi, z\}$.
For tensors and tensor-valued forms we sometimes suppress indices and use a
self-evident matrix notation (indicated by boldface types) instead.
Greek letters $\alpha, \beta, \dots=0,1,2,3$ will be used for anholonomic 
indices and Latin letters $i, j, \dots=0,1,2,3$ denote holonomic indices.
In case of doubt, anholonomic indices are marked by a hat.

The symbol $\wedge$ denotes the exterior product sign.  The frame (vector
basis) is named $e_\alpha=e^i{}_\alpha\partial_i$, the coframe
(one-form basis) $\omega^\beta=e_j{}^\beta \D x^j$; then we have
$e_\alpha\inner\omega^\beta=\delta^\beta_\alpha$, where $\inner$
denotes the interior product. The basis will be chosen {\em orthonormal}, 
i.e.\ $g=o_{\alpha\beta}\,\omega^\alpha\otimes\omega^\beta$ with
$o_{\alpha\beta}:={\rm diag}(-+++)$. Starting with the (metric) volume
four-form $\eta:=\omega^{\hat{0}}\wedge\omega^{\hat{1}}\wedge
\omega^{\hat{2}}\wedge\omega^{\hat{3}}$, we can successively
define the $\eta$-basis for forms according to $\eta_\alpha
:=e_\alpha \inner\eta $, $\eta_{\alpha\beta} :=e_\beta
\inner\eta_\alpha $, $\eta_{\alpha\beta\gamma} :=e_\gamma \inner
\eta_{\alpha\beta}$, $\eta_{\alpha\beta\gamma\delta} :=e_\delta \inner
\eta_{\alpha\beta\gamma} $. We will also make use of the short-hand
notation $e^\alpha := o^{\alpha\beta} e_\beta$ and $\omega_\alpha:=
o_{\alpha\beta}\,\omega^\beta$. The (Hodge) star operator will be
denoted by ${\,\!}^*$. We use geometrized units with $c=8\pi G=1$.

\section{Gauge structure: From Euclid to Poincar\'e}\label{Sect:Distortions}

A gravitational theory encompassing the equivalence principle is 
most conveniently described in terms of the geometry of spacetime. 
In this framework gravity is synonymous with some sort of deformation 
of spacetime.\footnote{The view that gravity is an expression of spacetime 
elasticity was adopted by 
Sakharov \cite{Sakharov67} who argued that the ``metric elasticity''
could be explained by quantum fluctuations of the vacuum. 
Here we follow the continuum mechanics analogy, but we do 
not pursue the search for a microscopic explanation of spacetime 
elasticity.}

Mass and spin are ``gravitational currents'', which in turn are intimately 
connected with the Poincar{\'e} group. From a gauge-theoretic perspective
one would a priori expect this group to be the local gauge group of gravity.  

Taking the elasticity picture seriously, it is of interest to systematically 
study simple elastic deformations of spacetime  and the corresponding
global geometries and local sources. To this end we generalize 
the Volterra process to four-dimensional spacetime.

\subsection{The generalized Volterra process}\label{Sec:VolterraProcess}
\begin{figure}[t]
     \setlength{\unitlength}{\textwidth}
\hbox to \psize{%
\raisebox{7mm}{{\epsfxsize = 0.25\psize\epsfbox{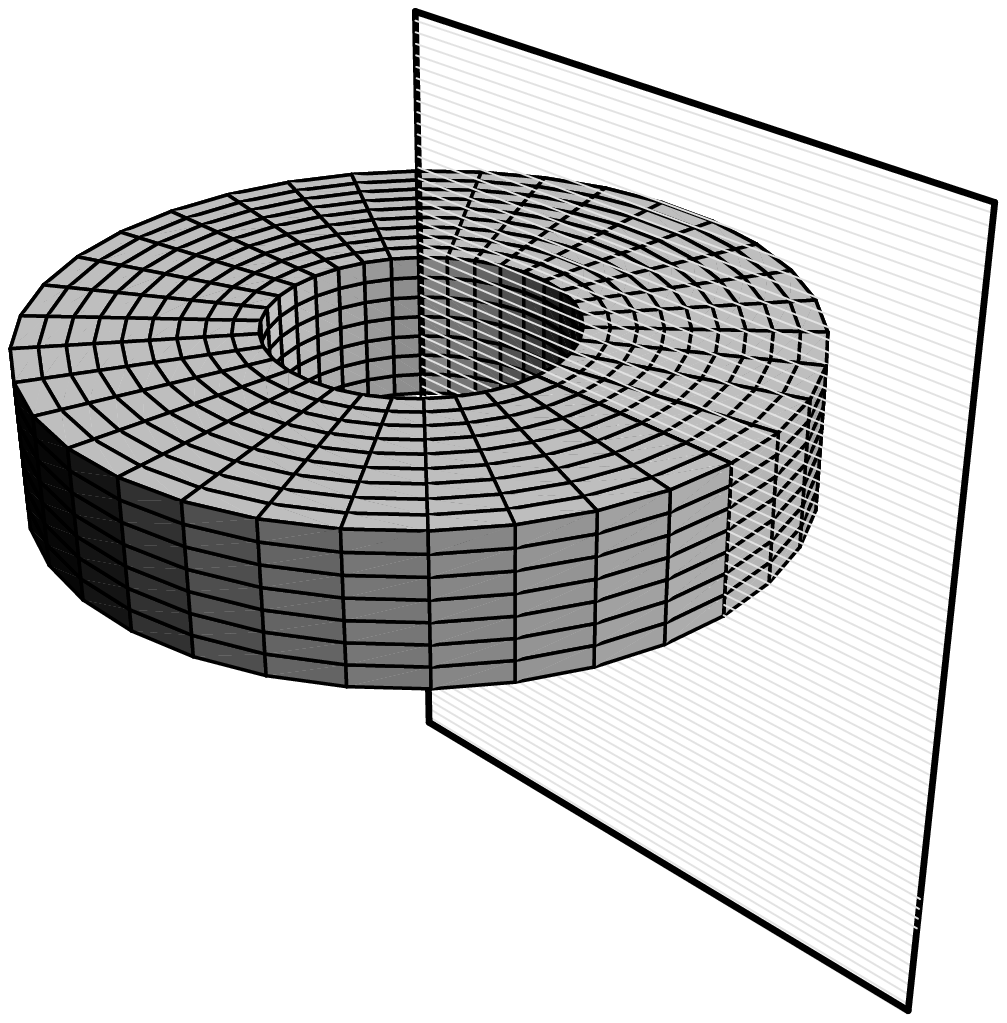}}}%
   \hfil
\raisebox{0cm}{{\epsfxsize = 0.25\psize\epsfbox{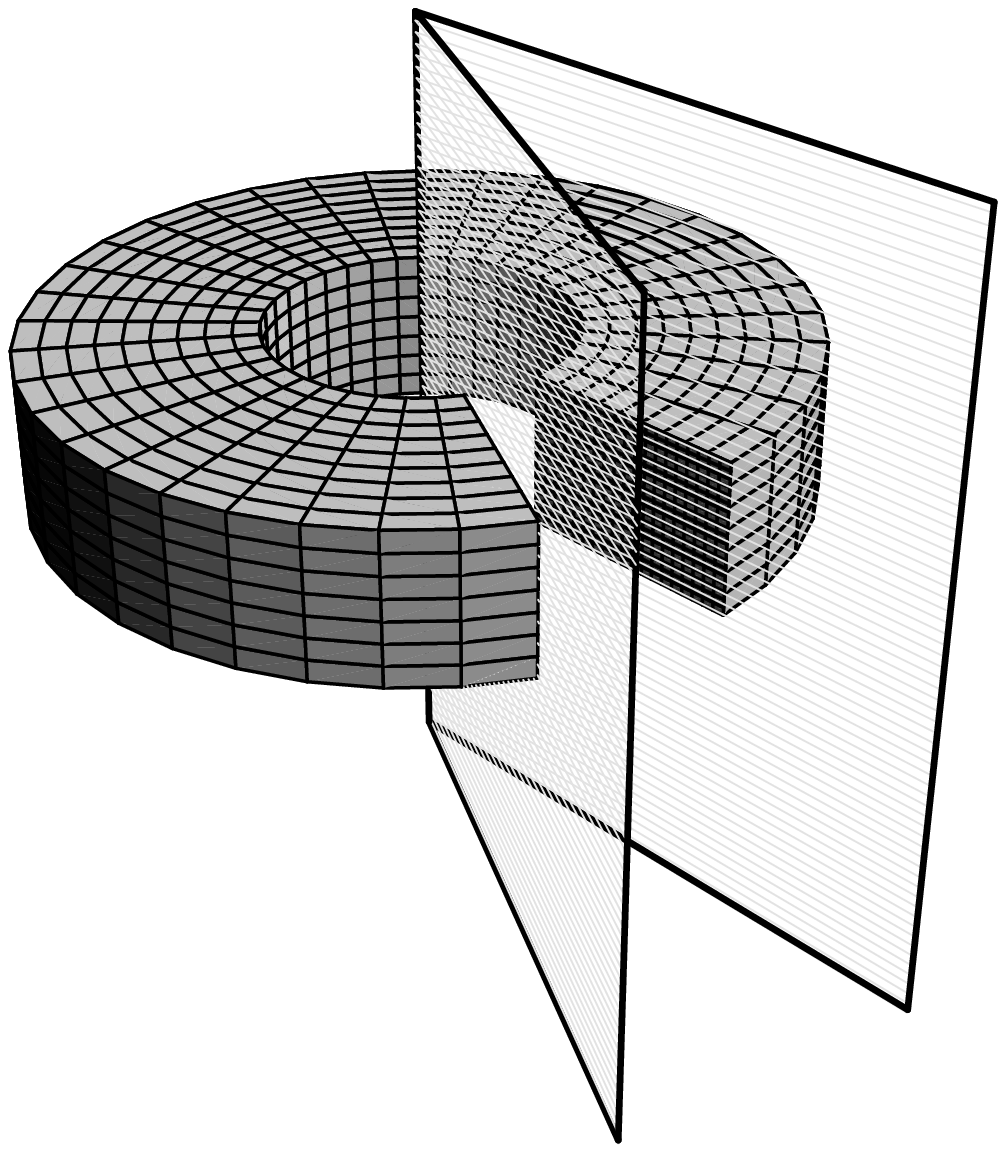}}}%
   \hfil
\raisebox{7mm}{{\epsfxsize = 0.29\psize\epsfbox{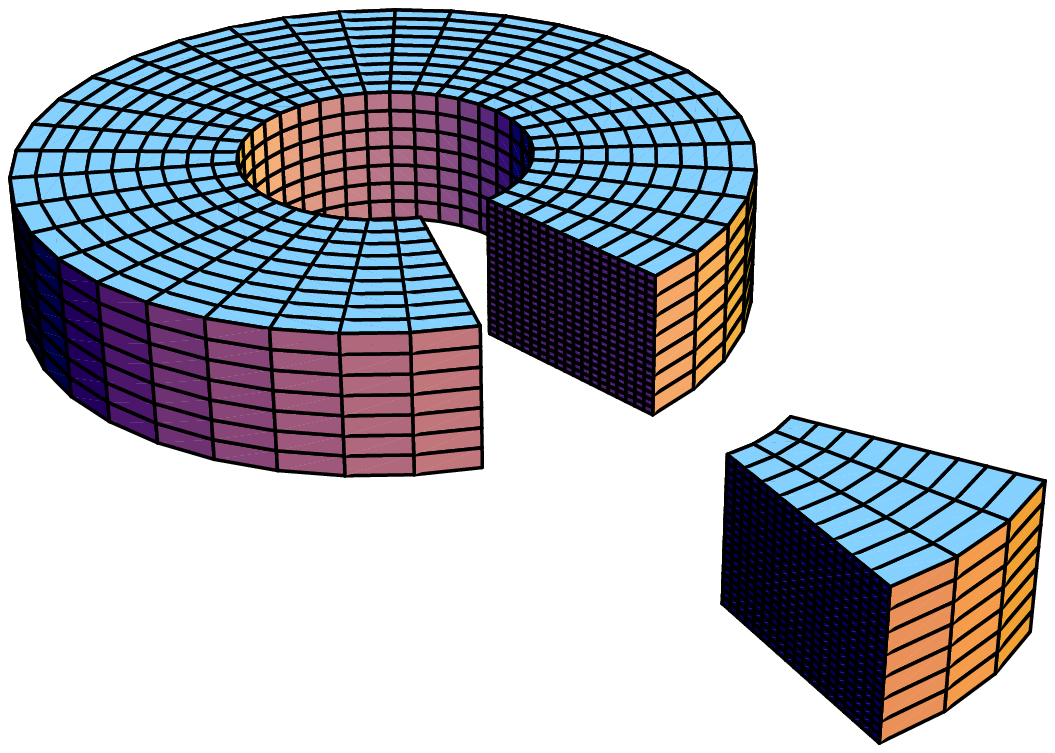}}}%
}

\vspace{-5cm}
\begin{picture}(1,.28)
   \put(.01,.0)  {(a)}
   \put(.38,.0)  {(b)}
   \put(.73,.0)  {(c)}
   \thicklines
   \put(0.575,0.017){\vector(-2,-1){.06}}
   \put(0.9,0.09){\vector(-3, 2){.047}}
\end{picture}
\caption{\label{fig:volterraProcess}The Volterra process for the wedge 
         disclination (order 6). (a) Cut hollow cylinder at a half two-plane.
         (b) Rotate lips. (c) Add missing material and weld 
         lips together}
\end{figure}

In the year 1907 Vito Volterra published an extensive article dedicated
to the study of elastic deformations of multiply-connected, solid, 
three-dimensional objects \cite{Volterra07}. 
His leitmotif was to take---as 
a prototype of such an object---a hollow cylinder made out of 
elastic material and cut it at a half two-plane, e.g.\ at $\phi = 0$ 
(using cylindrical coordinates $\{r, \phi, z\}$, the cylinder is taken 
to be oriented along the $z$-axis), thereby destroying
its multiple connectedness. Then take the two lips that have been
separated by the cut and translate and rotate them against each other.
Finally, after eventually removing superfluous or adding missing material, 
weld the two planes together again, see Fig.\ \ref{fig:volterraProcess}. 
This cutting and welding process is called the {\em Volterra process.}\par

\begin{figure}
     \setlength{\unitlength}{\textwidth}
\hbox to \psize{%
   \epsfxsize = 0.24\psize\epsfbox{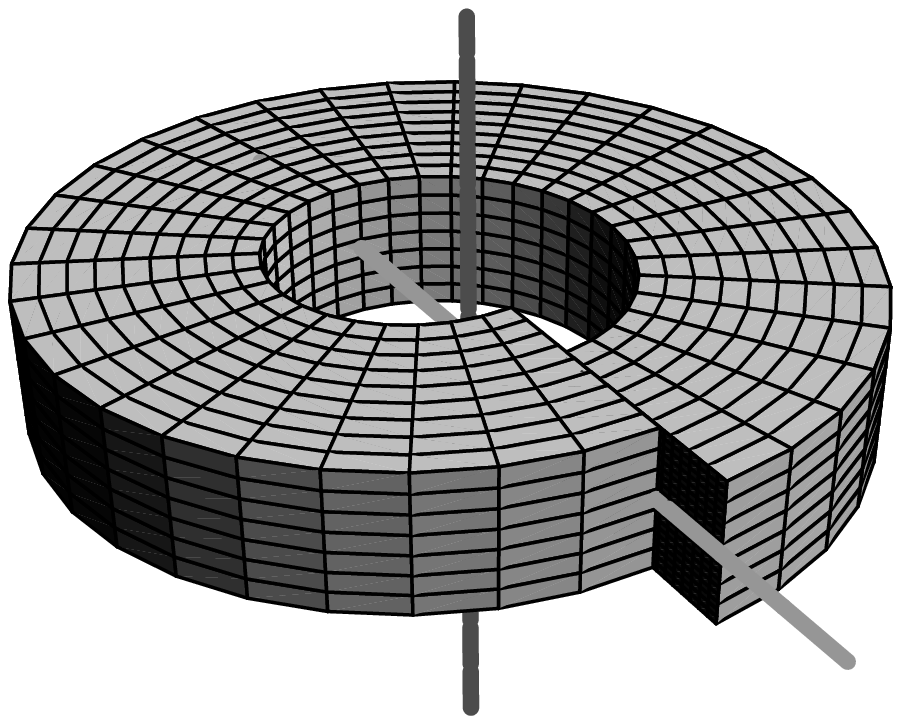}%
   \hfil
   \epsfxsize = 0.27\psize\epsfbox{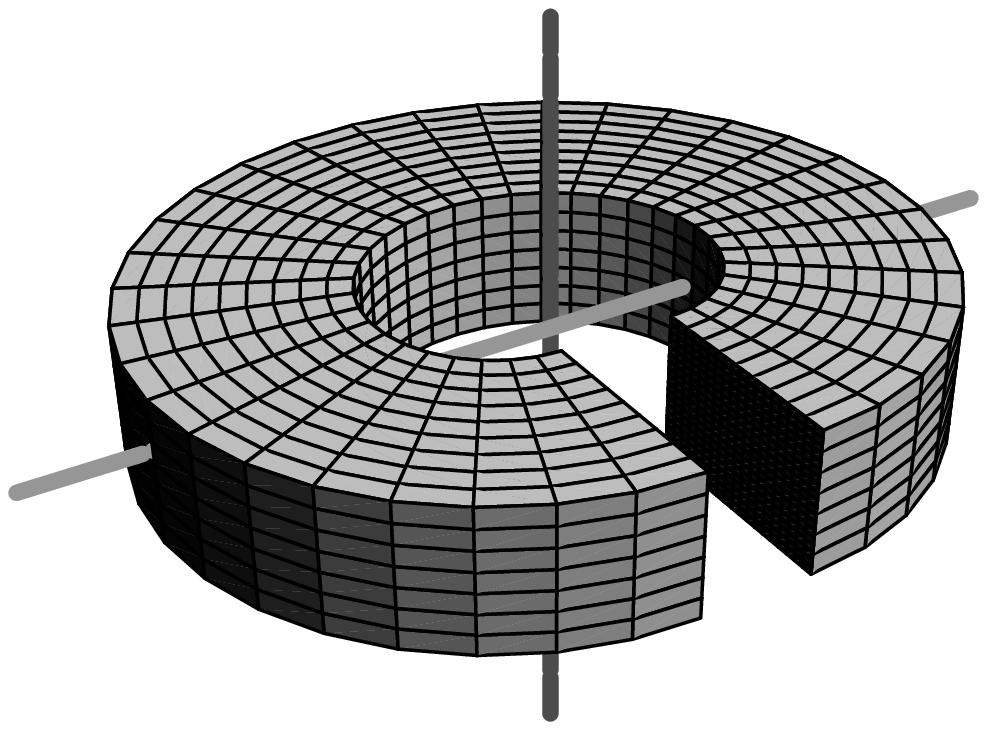}%
   \hfil
   \epsfxsize = 0.23\psize\epsfbox{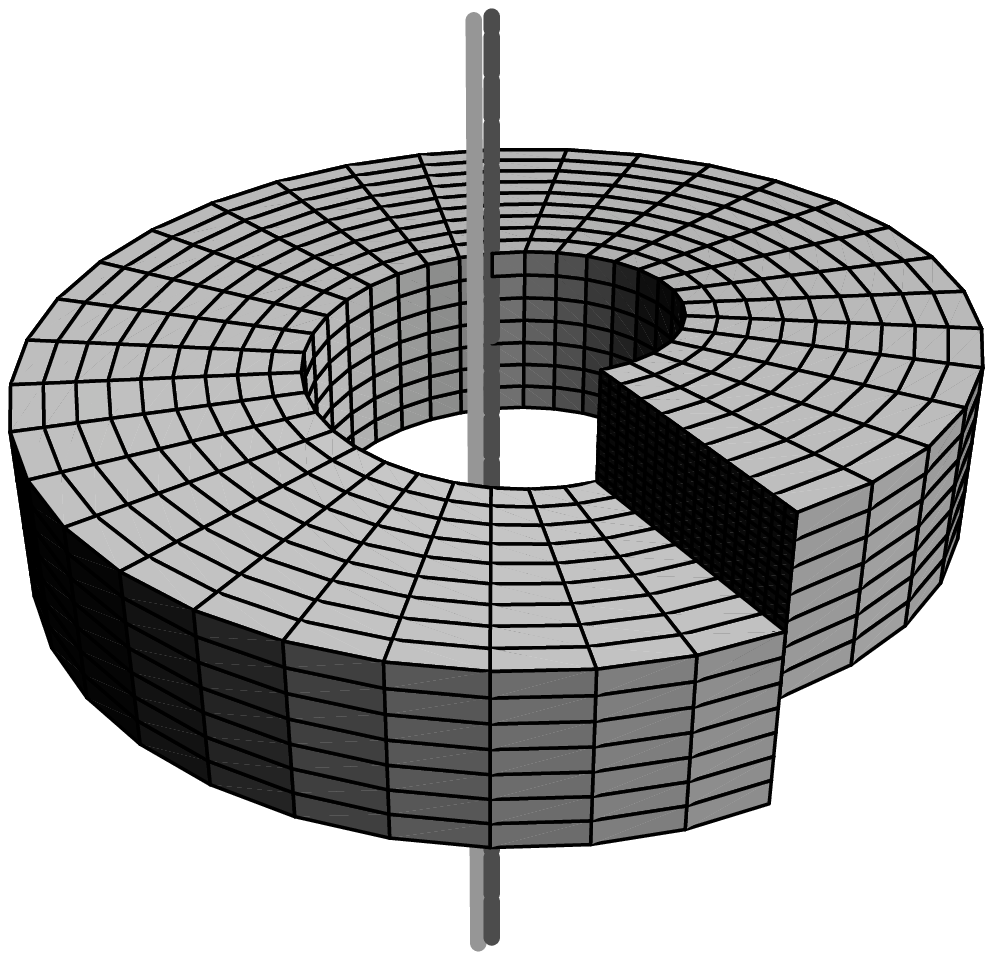}%
}
\smallskip
\hbox to \psize{%
   \epsfxsize = 0.24\psize\epsfbox{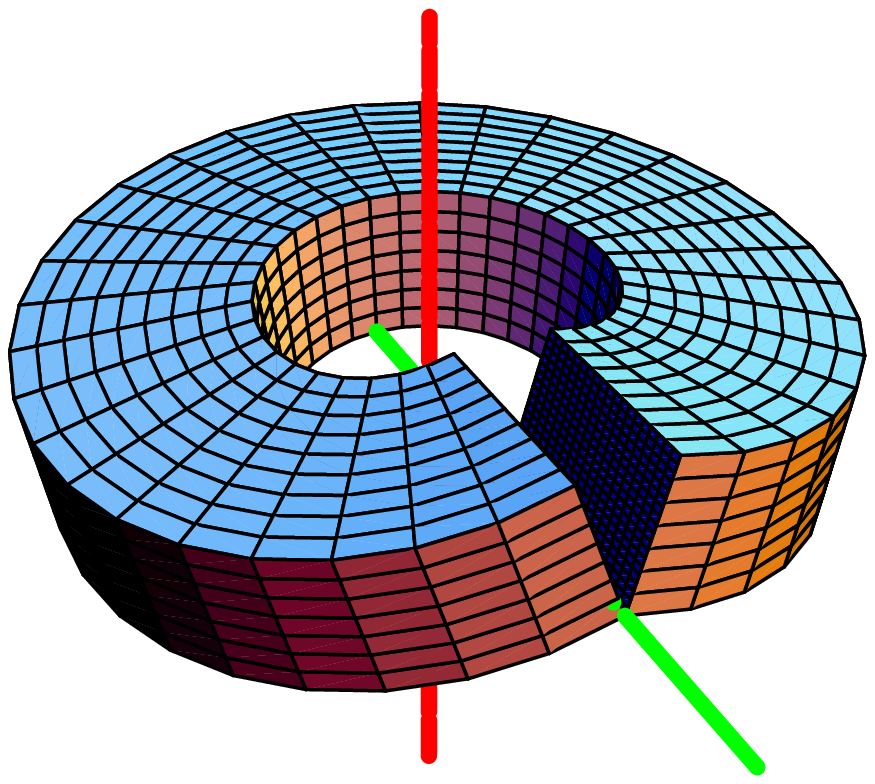}%
   \hfil
   \epsfxsize = 0.27\psize\epsfbox{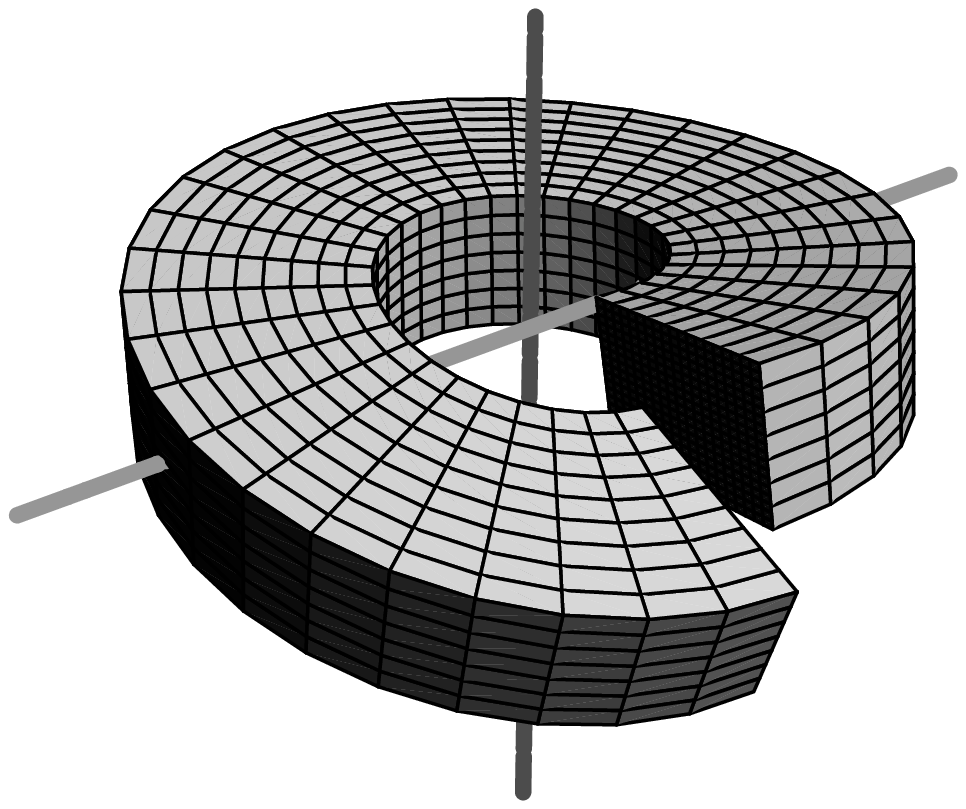}%
   \hfil
   \epsfxsize = 0.24\psize\epsfbox{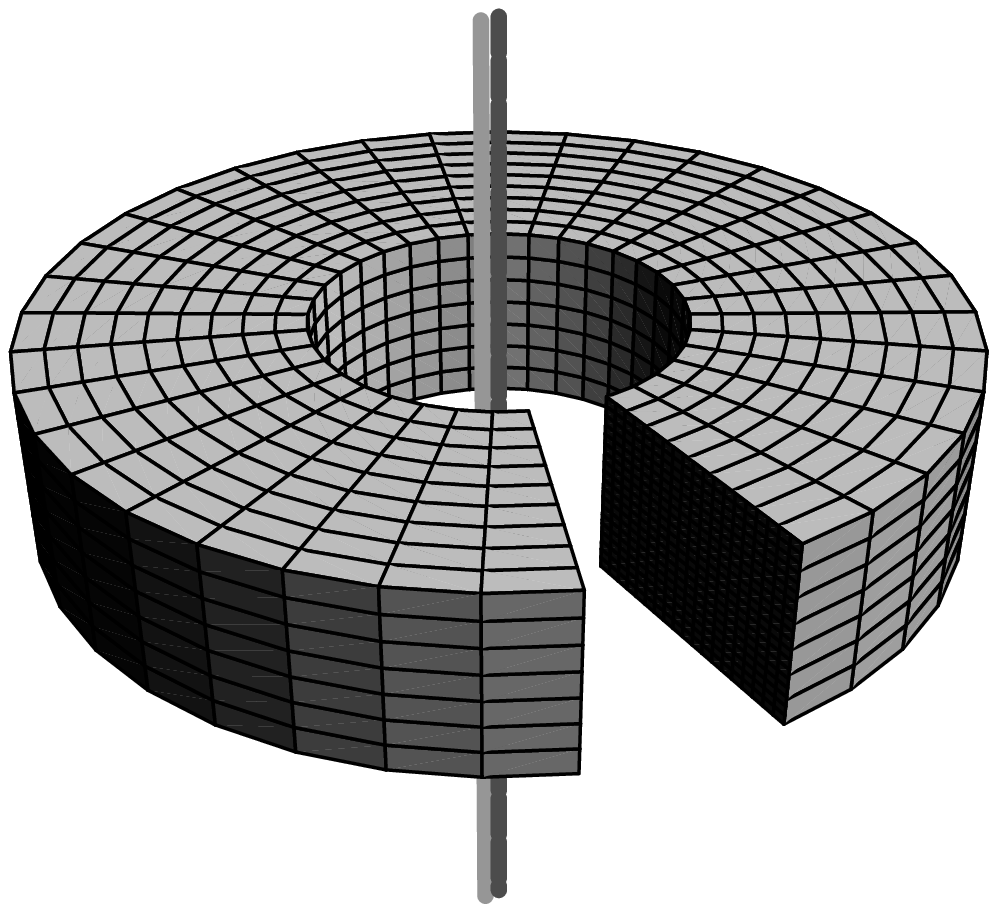}%
}
\vspace{-11.8cm}
\begin{picture}(1,.7)
   \put(.01,+.0)  {(d)}
   \put(.38,+.0)  {(e)}
   \put(.73,+.0)  {(f)}
   \put(.01,0.23){(a)}
   \put(.38,0.23){(b)}
   \put(.73,0.23){(c)}
\end{picture}
\caption{\label{fig:distortions}The six Volterra distortions. (a)--(c)
         Dislocations (order one to three). (d)--(f) Disclinations 
         (order four to six)} 
\end{figure}

The Volterra process, by construction, yields six different kinds of 
objects, see Fig.\ \ref{fig:distortions}, which belong to the six degrees 
of freedom of the proper group of motion in $\mathbb R^3$, the Euclidean 
group $SO(3) {\;{\rlap{$\supset$}\times}\;} T(3)$. Volterra called the 
resulting configurations  {\em distortions of order one to six}. The 
distortions belonging to the translational subgroup $T(3)$ and the rotational 
subgroup $SO(3)$ are called {\em dislocations\/} and {\em disclinations,}  
respectively.\par

When lifting the Volterra process from $3$ to $3 + 1$ dimensions, it is 
clear that instead of creating distorted hollow cylinders out of undistorted
ones, the basic notion must be to `distort' Minkowski spacetime into different
Riemann--Cartan geometries. The group that takes the place of 
$SO(3) {\;{\rlap{$\supset$}\times}\;} T(3)$ is evidently the Poincar\'e group 
$P(10) = SO(1,3) {\;{\rlap{$\supset$}\times}\;} T(4)$ with its 6 rotational 
and 4 translational parameters corresponding to the proper orthochronous 
Lorentz subgroup $SO(1,3)$ and the translational subgroup $T(4)$, 
respectively. Adopting the language of the theory of elasticity, 
we may say that Minkowski spacetime  will be deformed by 6 disclinations 
and 4 dislocations into 10 differently structured Riemann--Cartan 
spacetimes. It will turn out that these are locally flat and uncontorted, 
but contain---as expected---topological defects represented by singular
lines.\par 

To arrive at an accurate description of the Volterra process, we must
take a closer look at the cut that precedes the deformation.  In an
$n$-dimensional manifold, this cut is constructed in such a way that it
separates two half $(n-1)$-dimensional hyperspaces, the {\em lips\/}
of the cut, ending at an $(n-2)$-dimensional hyperspace, which will be
denoted as the {\em defect region}. This name is justified because the
defect region has to be excluded from the spacetime manifold,
in analogy to the interior region of Volterra's hollow cylinder.
The two-plane perpendicular to the defect will be called the {\em
  supporting plane\/} of the distortion.  It will carry translational
and rotational holonomy.  In a $(3+1)$-dimensional spacetime, there are
two situations, which are distinct in principle: the intrinsic metric
of the supporting plane may carry Riemannian (spacelike) or Lorentzian 
signature.  In the first case, the respective
distortions are said to be {\em space-supported}, in the latter case
they are called {\em space-\ and time-supported}. Of course, the space-\ 
and time-supported distortions have no analogy in the theory of
elasticity.

\subsection{Distorting spacetime by gauging the Poincar\'e group}
            \label{Sec:Voltmath}
In the theory of elasticity, the conventional approach to describe the 
mechanics of an elastic deformation is to displace the point 
\boldmath{$x$}\unboldmath\ of some unstressed body by an amount 
${\mbox{\boldmath{$u$}}} ({\mbox{\boldmath{$x$}}})$.  Thus the point 
with coordinates ${\mbox{\boldmath{$x$}}}'$ of the deformed body is
related to the initial state by
\[
   {\mbox{\boldmath{$x$}\unboldmath}}' =
   {\mbox{\boldmath{$x$}\unboldmath}} +
   {\mbox{\boldmath{$u$}\unboldmath}}
   ({\mbox{\boldmath{$x$}\unboldmath}})\;,
\]
and ${\mbox{\boldmath{$u$}}} ({\mbox{\boldmath{$x$}}})$ is called the
{\em displacement field\/} of the deformation. With the Volterra
process in mind, the deformation can be described more generally by saying that
the Euclidean group $SO(3) {\;{\rlap{$\supset$}\times}\;} T(3)$ acts {\em locally\/} on
the undeformed state, i.e.\ 
\begin{equation}
  {\mbox{\boldmath{$x$}\unboldmath}}' =
  {\mbox{\boldmath{${\mathcal{R}}$}\unboldmath}}
  ({\mbox{\boldmath{$x$}\unboldmath}})
  {\mbox{\boldmath{$x$}\unboldmath}} +
  {\mbox{\boldmath{${\mathcal{T}}$}\unboldmath}}
  ({\mbox{\boldmath{$x$}\unboldmath}}) \;,
\label{Eq:localEuclid}
\end{equation} 
where ${\mbox{\boldmath{${\mathcal{T}}$}\unboldmath}} \in T(3)$ and
${\mbox{\boldmath{${\mathcal{R}}$}\unboldmath}} \in SO(3)$.\par

Conventionally, the theory of elasticity is formulated by means of differential
geometric tools \cite{Kroner81,Kroner90,Mistura90}. The suggestive form of
(\ref{Eq:localEuclid})\ points to a more recent approach, namely a $SO(3)
{\;{\rlap{$\supset$}\times}\;} T(3)$ gauge theory of solid continua containing
defects \cite{Kadic83,Kroner86,Kleinert89,Rivier90,Malyshev93}, which
will allow the generalizations envisioned very intuitively (a corresponding 
task in the context of $2+1$ dimensional gravity is investigated in the work 
by Kohler \cite{Kohler95,Kohler95b}): we can, in a
natural way, adapt the results from elasticity to a theory of gravity
by taking the Poincar\'e group $P(10)$ as a gauge group acting locally on 
Minkowski spacetime $M_4$. In essence, we are developing an {\em active\/} 
Poincar\'e gauge theory . The reason why we select the
Poincar\'e group is its prominent r{\^o}le as a spacetime symmetry group
in special relativity. Moreover, both mass and spin, the corresponding 
matter currents, are known to exist in nature.  

Consequently, our starting point is an
$M_4$ with global coordinate 
cover $\mbold x = \{x^0,x^1,x^2,x^3\}$.
In analogy to (\ref{Eq:localEuclid}), we express the
local action of the
Poincar\'e group $P(10)$ by the operation
\begin{equation}
  {\mbox{\boldmath{$x$}\unboldmath}}' =
  \mcal P (\mbold x) \, \mbold x :=
  {\mbox{\boldmath{${\mathcal{L}}$}\unboldmath}}
  ({\mbox{\boldmath{$x$}\unboldmath}})
  {\mbox{\boldmath{$x$}\unboldmath}} +
  {\mbox{\boldmath{${\mathcal{T}}$}\unboldmath}}
  ({\mbox{\boldmath{$x$}\unboldmath}})
\label{Eq:x'}
\end{equation} 
with a local Lorentz boost $\mcal L (\mbold x)$ and a local translation
$\mcal T (\mbold x)$. The gauge construction will become most 
transparent if we introduce the following M\"obius type matrix 
representation \cite{Tseytlin82,Ivanenko83,Mielke93a,Hehl95}: in the 
five-dimensional 
hyperplane $\tilde{\tilde{\mathbb R^4}}:= \left\{ \tilde{\tilde{\mbold x}} 
 := (\mbold x, 1)^t \in \mathbb R^5 \right\}$, the
group action (\ref{Eq:x'}) can be written
\begin{equation}
   \tilde{\tilde{\mbold x}}' = \mcal P\, \tilde{\tilde{\mbold x}} =   
   \left(\begin{array}{c c}
            \mcal L & \mcal T \\
                  0 &       1
            \end{array}
   \right) 
   \left(\begin{array}{c}
            \mbold x\\
                   1
         \end{array}
   \right) =
   \left(\begin{array}{c}
            \mcal L \mbold x  + \mcal T \\ 
                                      1
         \end{array}
   \right)\;.
\end{equation}
Here and in the rest of the article it is understood that operations 
always act locally, i.e.\ $\mcal L = \mcal L (\mbold x )$, and so on. 
Now the standard gauge approach \`a la  Yang--Mills can be 
applied \cite{Mielke87}. We define the connection 
\begin{equation}
    \tilde{\tilde{\mbold \Gamma}} = 
     \left(
     \begin{array}{c c}
      \mbold \Gamma^{({\rm L})} & \mbold \Gamma^{({\rm T})}\\
       0                        & 0
      \end{array}\right) 
\end{equation}
that transforms inhomogeneously
into
\begin{equation}
{\tilde{\tilde{\mbold \Gamma}}}' = \mcal P 
          \tilde{\tilde{\mbold \Gamma}} \mcal P^{-1}
          - \D  \mcal P \, \mcal P^{-1}
\end{equation}
under $P(10)$. Explicitly, we find the $SO(1,3)$-algebra-valued 
connection one-form, the {\em Lorentz connection},
\begin{equation}
  {\mbold \Gamma^{({\rm L})}}' = 
                   \mcal L \mbold \Gamma^{({\rm L})} {\mcal L}^{-1}
                             - (\D \mcal L) {\mcal L}^{-1} \;,
\label{Eq:connection}
\end{equation}
compensating for the local action of the Lorentz sector, and
the $\mathbb R^4$-valued connection one-form, the {\em translation 
connection},
\begin{equation}
   {\mbold \Gamma^{({\rm T})}}' 
        =  \mcal L \mbold \Gamma^{({\rm T})} - \DD\mcal T
        =   \mcal L \mbold \Gamma^{({\rm T})} - \D \mcal T
                - {\mbold \Gamma^{({\rm L})}}' \, \mcal T  \;,
\label{Eq:Phi}
\end{equation}
compensating for the local action of $T(4)$. Here we have used the
notation $\DD= \D + {\mbold \Gamma^{({\rm L})}}' \wedge$.

The transformation behaviour (\ref{Eq:Phi}) of the translation part
$\mbold \Gamma^{({\rm T})}$ of the connection is inhomogeneous and hence,
in particular, nontensorial. For this reason, $\mbold \Gamma^{({\rm T})}$
cannot be taken as coframe of the target space. There is, however, a trick
to circumvent this situation \cite{Hehl95}: the {\em soldering one-form\/}
\begin{equation}
  \mbold \omega  := {\mbold \Gamma^{({\rm T})}}' + \DD\, \mbold\xi
                 =  {\mbold \Gamma^{({\rm T})}}' + 
                   \D\mbold\xi + {\mbold \Gamma^{({\rm L})}}'\mbold\xi\;,
\end{equation}
with some vector-valued zero-form $\mbold\xi$, shows the `right' 
transformation behaviour, namely that of a vector-valued one-form.
The r\^ole of $\mbold\xi$, sometimes called the
{\em Poincar\'e 
coordinate\/} \cite{Grignani92}, or {\em generalized Higgs field\/} 
\cite{Trautman79}, is not completely clear, see \cite{Hehl95} and the 
literature given there. In contrast to the authors of \cite{Hehl95}, who 
choose $\DD\, \mbold\xi = 0$, we impose the condition $\mbold\xi = \mbold x$,
and find
\begin{equation}\label{Eq:coframe}
  \mbold \omega  = \mbold \Gamma^{({\rm T})} + \DD\, \mbold x
                 = \mbold \Gamma^{({\rm T})} + 
                   \D\mbold x + {\mbold \Gamma^{({\rm L})}}\mbold x\;,
\end{equation}
which can be regarded as the exterior covariant derivative 
$\tilde{\tilde \DD} := \D + \tilde{\tilde \mbold \Gamma} \wedge$ 
acting on $\tilde{\tilde \mbold x}$:
\begin{equation}
     \tilde{\tilde \mbold \omega} :=  
     \tilde{\tilde \DD}\, \tilde{\tilde \mbold x} = 
     \left(
     \begin{array}{c}  
      \mbold \Gamma^{({\rm T})} +
      \D\mbold x + {\mbold \Gamma^{({\rm L})}}\mbold x\\ 
       0
      \end{array}\right)\;. 
\end{equation}
The soldering forms (\ref{Eq:coframe}) make up
an orthonormal coframe, 
which yields 
\begin{equation}\label{eq:lineElement}
   g = g_{ij}\, \D x^i \otimes \D x^j
     := o_{\alpha\beta}\,\omega^\alpha\otimes\omega^\beta \;,
\end{equation}
with $o_{\alpha\beta}:={\rm diag}(-+++)$. 
Finally, the torsion two-form \boldmath{$T$}\unboldmath\ is defined by the
structure equation
\begin{equation}\label{eq:structure1}
  \mbold T = \D\, \mbold\omega + \mbold\Gamma^{({\rm L})} \wedge \mbold\omega 
       = \mbold {R x} + \D\, \mbold \Gamma^{({\rm T})} 
         + \mbold\Gamma^{({\rm L})} \wedge \mbold\Gamma^{({\rm T})} \;,
\end{equation}
where \boldmath{$R$}\unboldmath\ denotes the 
curvature two-form defined by the structure equation
\begin{equation}\label{eq:structure2}
   \mbold R = \D \mbold\Gamma^{({\rm L})}
            + \mbold\Gamma^{({\rm L})} \wedge \mbold\Gamma ^{({\rm L})} \;.
\end{equation}
\subsection{Burgers vector and Frank matrix}
We have already pointed to the fact that the distorted spacetime manifolds 
will be {\em locally\/} flat and uncontorted. In order to 
detect the effect of the generalized Volterra process, we compute the 
holonomy transformation, usually considered in the loop space formulation
of gauge theories \cite{Yang74}. (See also \cite{Bezerra91} in the context 
of cosmic strings.) 

For our purpose, it is sufficient to think of a vierbein, transported
along a closed path $S$ around the line-like defect region. More precisely,
$S$ is the image in $U_4$ of a closed path\footnote{Very similarly, 
we can take $S$ to be the {\em development\/} \cite{Petti86} in $M_4$ of a 
curve which  is closed in $U_4$.} in the undistorted $M_4$ under the 
mapping (\ref{Eq:x'}).
As a result of this parallel transport, we will detect
translational holonomy,  characterized by  the {\em Burgers vector\/} 
${\mbox{\boldmath{${\mathcal{B}}$}}} \in T(4)$, 
and rotational holonomy, described by 
${\mbox{\boldmath{${\mathcal{G}}$}\unboldmath}} \in SO(1,3)$,
the {\em Frank matrix}.\footnote{G\"ockeler 
and Sch\"ucker \cite{Gockeler87}
use the name {\em parallel  transporter\/} for 
${\mbox{\boldmath{${\mathcal{G}}$}\unboldmath}}$. However, to be 
consistent we would have to use something like ``translational parallel 
transporter'' for the Burgers vector. Therefore, we have adapted the name
Frank  matrix, which is the four-dimensional analogue of the 
Frank vector of a solid-state disclination.} 
In other words, after parallel transport along $S$, the vierbein will have
been subject to an affine transformation---the above-mentioned holonomy 
transformation---namely a translation \boldmath{${\mathcal{B}}$}\unboldmath\
and a ``rotation'' ${\mbox{\boldmath{${\mathcal{G}}$}\unboldmath}}$.

The Burgers vector is defined to be the contour integral of the 
coframe \cite[p.~1372]{Kleinert89}
\[
    {\mbox{\boldmath{${\mathcal{B}}$}\unboldmath}} 
     \equiv \oint_S {\mbox{\boldmath{$\omega$}\unboldmath}}\;. 
\]
The contour integral of the Minkowski coframe  
$\D \mbold x$ vanishes. Thus, if we restrict ourselves
to dislocations, characterized by $\mcal L = \mbold 1$ and consequently 
$\mbold \Gamma^{({\rm L})} = 0$, we find from (\ref{Eq:coframe})
for the Burgers vector
\begin{equation}\label{Eq:BurgersVector}
    {\mbox{\boldmath{${\mathcal{B}}$}\unboldmath}} 
       = \oint_S \mbold \Gamma^{({\rm T})}
       = -\oint_S \D \mcal T \;.
\end{equation}
In order to compute the Frank matrix 
${\mbox{\boldmath{${\mathcal{G}}$}\unboldmath}}$ 
of the distorted spacetime, 
we start from the Lorentz connection $\mbold \Gamma^{({\rm L})}$
as defined in (\ref{Eq:connection}). The Frank matrix is defined 
as the Lorentz connection integrated along the path $S$,
resulting in an element of the Lorentz group. Explicitly we have
\begin{equation}\label{Eq:FrankMatrix}
   {\mbox{\boldmath{${\mathcal{G}}$}\unboldmath}} = 
     {\mathcal P}\exp \left\{-\oint_S \mbold \Gamma^{({\rm L})}
                      \right\}\;,
\end{equation}
where ${\mathcal{P}}\exp$ denotes the path-ordered exponential, which reflects
the noncommutative structure of the Lorentz group $SO(1,3)$. Full information
on the rotational holonomy is only contained in the Frank matrix. 
The deficit angle $\Delta\phi$, frequently used to describe
disclinations in solids, is the fixed parameter value of 
${\mbox{\boldmath{${\mathcal{G}}$}\unboldmath}}$, but, 
of course, we lose information on the two-plane in which 
${\mbox{\boldmath{${\mathcal{G}}$}\unboldmath}}$ 
acts if we only give the deficit angle $\Delta\phi$.

\section{Volterra distortions}\label{s:volterra}
\begin{table} 
\caption{\protect\label{tab:disloc+disclin}
         Dislocations and disclinations characterized by the geometric objects
         that are introduced in Section \protect\ref{Sect:Distortions}.}
\medskip
\begin{tabular}{l  l  l}
\hline
& Dislocation of order $i$ & Disclination of order $a$\\
\hline
Rotation&
$\mcal L  = \mbold 1$           & 
                             $\mcal L = \exp\left(\rule[-2pt]{0mm}{4.5mm}
                                  \Phi/(2 \pi) \phi\, \mbold l_a\right)$\\
Translation&
$\mcal T = J/(2\pi) \phi\, \mbold e_i$ 
                                & $\mcal T  = \mbold 0$ \\
Lorentz connection&
$\mbold \Gamma^{({\rm L})} = 0$ & $\mbold \Gamma^{({\rm L})} = 
                                  -\D \mcal L  \mcal L^{-1}$ \\
Translation connection&
$\mbold \Gamma^{({\rm T})} = - \D \mcal T$ 
                                & $\mbold \Gamma^{({\rm T})} = 0$\\
Soldering one-form&
$\mbold \omega = \D\mbold x - \D \mcal T$
                                & $\mbold \omega =  \D \mbold{x} 
                                   + \mbold \Gamma^{({\rm L})}\mbold x
                                   =  \D \mbold x 
                                   - \D \mcal L \mcal L^{-1}\mbold x$\\
Burgers vector&
$\mcal B = \displaystyle\oint \left( \D \mbold x - \D \mcal T \right)
=  \displaystyle\frac{J}{2 \pi}\, \mbold e_i$ 
                              & $\mcal B = \displaystyle\oint\left(\D \mbold x
                            - \D \mcal L \mcal L^{-1}\,\mbold x\right)$\\
Frank matrix&
$\mcal G = 0$                   & $\mcal G = {\cal P}\exp \displaystyle\oint  
                                  -\D \mcal L  \mcal L^{-1}$\\
\hline
\end{tabular}
\end{table}
Before we can explicitly construct the Volterra distorted spacetimes, we
must specify the geometry by picking the two-plane supporting the
defect. In the present section, we study space-supported distortions and,
using Cartesian coordinates $\mbold x = \{x^0,x^1,x^2,x^3\} = \{t,x,y,z\}$ 
for the undistorted $M_4$, we choose the $x$--$y$ plane, 
conveniently 
parametrized 
by the radius $r = \sqrt{x^2 + y^2}$ and the polar 
angle $\phi = \arctan y/x$.

As we are assuming Cartesian coordinates for $M_4$, we have  
$\mbold \Gamma^{({\rm L})} = \mbold \Gamma^{({\rm T})} = 0$ in 
(\ref{Eq:connection}) and (\ref{Eq:Phi}). Therefore we can
omit primes for the transformed gauge connections, implying that whenever
connection one-forms $\mbold \Gamma^{({\rm L})}$ or $\mbold \Gamma^{({\rm T})}$
show up, they belong to the {\em distorted\/} spacetime $U_4$.

The notion that leads to the fundamental Volterra distortions is the following.
{\em Dislocations}, i.e.\ translational distortions, are defined by
$\mcal L = \mbold 1$ and $\mcal T = J/(2 \pi) \phi\, \mbold e_i$,
where $\{\mbold e_i \,|\, 0\leq i \leq 3\}$ are the generators of the 
translation group. The constant $J$ parametrizes the dislocation strength.
Using Volterra's original terminology, we may say that $\mcal T$, if defined
in such a way, generates a dislocation {\em of order\/} $i$.
Evaluating (\ref{Eq:connection}), we find that the Lorentz
connection one-form vanishes identically,
$\mbold \Gamma^{({\rm L})} = \mbold 0$, i.e.\ spacetime remains globally 
flat. From (\ref{Eq:Phi}) the translation connection is 
$\mbold \Gamma^{({\rm T})} = - \D \mcal T$
and the orthonormal coframe (\ref{Eq:coframe})\ evaluates to
\begin{equation}\label{Eq:Bi_t}
   \mbold \omega = \D \mbold x + \mbold \Gamma^{({\rm T})} 
                 = \D \mbold x - \D \mcal T \;.
\end{equation} 
Analogously we define rotational distortions or {\em disclinations}. Here we 
have $\mcal T = 0$ and $\mcal L = \exp\left(\rule[-1pt]{0mm}{4.1mm} 
\Phi/(2 \pi) \phi\,\mbold l_a\right)$,
with constant $\Phi$, the disclination strength, and the generators 
$\{\mbold l_a \,|\, 4\leq a \leq 9\}$ of the Lorentz group $SO(1,3)$. The 
boost $\mcal L$ belongs to a disclination {\em of order\/} $a$.
The translation connection $\mbold \Gamma^{({\rm T})}$ vanishes and 
(\ref{Eq:connection})\ evaluates to 
\begin{equation}
   \mbold \Gamma^{({\rm L})} = - \D \mcal L\, \mcal L^{-1} \;.
\label{Eq:DisclConnection}
\end{equation}
We have collected the geometric objects characterizing all space-supported 
dislocations and disclinations in Table \ref{tab:disloc+disclin}. Note that 
our definition of the order of the Volterra distortions is such that orders
one to six correspond to the classical Volterra distortions \cite{Volterra07}.
Order zero, the time dislocation, and orders seven to nine, the boost 
disclinations, are due to the generalization to spacetime. 

\subsection{Dislocations}
\begin{table} 
\caption{\protect\label{tab:disloc}
        Line elements for the space-supported dislocations}
\medskip
\begin{tabular}{l l}
\hline
Order\quad & Line element\\
\hline
0 & $ \D s^2 =  -\left(\D t + \frac{\Theta^0}{2\pi}\D\phi\right)^2
              + \D r^2 
              + r^2 \D\phi^2 
              + \D z^2$\rule{0mm}{7mm} \\[1mm]
1 & $ \D s^2 =  - \D t^2 + \D x^2 + \D y^2 + \D z^2 
          + 2\frac{\Theta^1}{2 \pi\;r^2}\; \D x \;(x\; \D y - y\; \D x)
          + \left(\frac{\Theta^1}{2\pi\;r^2}\right)^2(x\; \D y - y\; \D x)^2$\\[1mm]
2 & $ \D s^2 =  - \D t^2 + \D x^2 + \D y^2 + \D z^2
          + 2\frac{\Theta^2}{2 \pi\;r^2}\; \D y \;(x\; \D y - y\; \D x)
          + \left(\frac{\Theta^2}{2\pi\;r^2}\right)^2(x\; \D y - y\; \D x)^2$\\[1mm]
3 & $ \D s^2 = -\D t^2 + \D r^2 + r^2 \D\phi^2 
          + \left(\D z + \frac{\Theta^3}{2\pi}\D\phi \right)^2$ \\[1mm]
\hline
\end{tabular}
\end{table}
With Table \ref{tab:disloc+disclin} and formula (\ref{eq:lineElement}), 
the  line elements of the four 
space-supported Volterra dislocations are computed straightforwardly. 
The result is given in Table \ref{tab:disloc}.  
If the Burgers vector is contained in the defect region, the defect
corresponds to a {\em screw dislocation}, otherwise, i.e.\ the Burgers vector 
is perpendicular to the defect region, it is referred to as an
{\em edge 
dislocation}. Thus we have two screw dislocations (orders zero and three) and
two edge dislocations (orders one and two). The time dislocation 
(order zero) is 
due to the generalization to spacetime.

By using (\ref{eq:structure1}), we can compute the torsion two-forms of 
the four Volterra dislocations. If we take, for instance, the time dislocation 
(order zero), we find 
\begin{eqnarray}
   T^0 & = &\D \omega^0  =  -\frac{\Theta^0}{2\pi}\, \D 
         \left(\partial_x\phi\, \D x  + 
               \partial_y\phi\, \D y\right)
        =  -\frac{\Theta^0}{2\pi} 
         \left(-\partial_y\partial_x\,\phi + 
                \partial_x\partial_y\,\phi
         \right) \, \D x \wedge \D y \nonumber\\
        & = & -\Theta^0 \,\delta^2(x,y) \, \D x \wedge \D y \;, 
             \label{eq:torsion0} 
\end{eqnarray}
where $\delta^2(x,y)$ is the two-dimensional delta distribution 
supported by the $x$--$y$ plane. Here, we have basically used the identity 
$\D^2 \phi \equiv \D^2 \arctan y/x = 2\pi\delta^2(x,y)\, \D x \wedge \D y $.
All other torsion components vanish. The corresponding computation 
for the remaining three dislocations is done in complete analogy.
\subsection{Disclinations}\label{Sec:Disclination}
Let us investigate, as the prototype of a disclination, some details of
the axial {\em wedge\/} disclination (order six). It is characterized by 
the local rotation
\begin{equation}
    {\mbox{\boldmath{${\mathcal{L}}$}\unboldmath}} = \left(
    \begin{array}{c c c c}
      1 &                       0 &                         0 & 0 \\
      0 & \cos\left(\frac{\Phi^6}{2\pi}\phi\right) 
        & -\sin\left(\frac{\Phi^6}{2\pi}\phi\right) & 0 \\
      0 & \sin\left(\frac{\Phi^6}{2\pi}\phi\right) 
        &  \cos\left(\frac{\Phi^6}{2\pi}\phi\right) & 0 \\
      0 &                       0 &                         0 & 1 
    \end{array}\right)
\end{equation}
in the $x$--$y$ plane.
After some elementary algebra, (\ref{Eq:DisclConnection})\ yields 
\begin{equation}
   {\mbold \Gamma^{({\rm L})}} = \left(
    \begin{array}{c c c c}
        0 & 0 &  0 & 0 \\
        0 & 0 &  1 & 0 \\
        0 &-1 &  0 & 0 \\
        0 & 0 &  0 & 0 
    \end{array}\right) \frac{\Phi^6}{2 \pi r^2}
       (x\; \D y - y\;\D x)
\end{equation}
for the Lorentz connection. By using (\ref{Eq:FrankMatrix}), we compute the 
Frank matrix of an arbitrary loop $S$,
\begin{equation} 
   {\mbox{\boldmath{${\mathcal{G}}$}\unboldmath}} = 
    {\mathcal P}\exp \left\{-\oint_{\phi = 0}^{2 \pi} 
     \! {\mbox{\boldmath{$\Gamma$}\unboldmath}}^{({\rm L})} \right\} =
    \left(
    \begin{array}{c c c c}
      1 &            0 &             0 & 0 \\
      0 &  \cos \Phi^6 &   \sin \Phi^6 & 0 \\
      0 & -\sin \Phi^6 &   \cos \Phi^6 & 0 \\
      0 &            0 &             0 & 1 
    \end{array}\right) \;,
\label{G(6)}
\end{equation} 
corresponding to the deficit angle $\Delta\phi = \Phi^6$.
These results are very intuitive if one remembers the 
usual interpretation
of the geometric objects involved: the connection 
$\mbold\Gamma^{({\rm L})}$ is a
Lorentz-{\em algebra\/}-valued one-form describing {\em infinitesimal\/}
parallel transport of a vector frame. The Frank 
matrix ${\mbox{\boldmath{${\mathcal{G}}$}\unboldmath$$}}$ 
takes its values in the Lorentz {\em group\/} corresponding to parallel 
transport of a vector frame along the finite path $S$. 

The orthonormal coframe (\ref{Eq:coframe})\ becomes 
\begin{equation}
\left. \begin{array}{l}\displaystyle 
   \omega^0  =  \D  t \\
   \omega^1  =  \D x + \frac{\Phi^6}{2\pi}\frac{y}{r^2} \,(x \D y - y \D x) \\
   \omega^2  =  \D y - \frac{\Phi^6}{2\pi}\frac{x}{r^2} \,(x \D y - y \D x) \\
   \omega^3  =  \D z \;.
      \end{array}\right.
\end{equation}
The curvature two-form may be  calculated  from the structure 
equation (\ref{eq:structure2}). With 
the Lorentz connection (\ref{G(6)}) and a computation similar to the 
one resulting in (\ref{eq:torsion0}), the nonvanishing components of 
the curvature two-form turn out to be
\begin{equation}\label{eq:curvOrder6}
   {R^1}_2 = - {R^2}_1 = \Phi^6\,\delta^2(x,y) \,\D x \wedge \D y \;.
\end{equation}
The curvature of the spacetimes corresponding to the remaining 
disclinations is calculated straightforwardly; the result is obvious and 
therefore not given explicitly.
The line elements corresponding to the six Volterra disclinations are 
given in Table \ref{tab:disclin}. 
\begin{table}
\caption{\protect\label{tab:disclin}
        Line elements for the space-supported disclinations}
\medskip
\begin{tabular}{l l}
\hline
Order\quad & Line element\\
\hline
4 & $ \D s^2 =  - \D t^2 + \D x^2 + \D y^2 + \D z^2$\\
  &           \phantom{$ \D s^2 =  $} $
              + 2\frac{\Phi^4}{2 \pi\;r^2}(z\; \D y - y\; \D z)\;
              (x\; \D y - y\; \D x)
              + \left(\frac{\Phi^4}{2\pi\;r^2}\right)^2
                \left(y^2 + z^2\right)(x\; \D y - y\; \D x)^2$ \\[1mm]
5 & $ \D s^2 =  - \D t^2 + \D x^2 + \D y^2 + \D z^2$\\
  &           \phantom{$ \D s^2 =  $} $
              + 2\frac{\Phi^5}{2 \pi\;r^2}(z\; \D x - x\; \D z)\;
              (x\; \D y - y\; \D x)
              +  \left(\frac{\Phi^5}{2\pi\;r^2}\right)^2
                \left(x^2 + z^2\right)(x\; \D y - y\; \D x)^2$\\[1mm]
6 & $ \D s^2 =  -\D t^2 + \D r^2
              +\left(1 - \frac{\Phi^6}{2\pi}\right)^2 r^2 \D\phi^2 
              + \D z^2$\\[1mm]
7 & $ \D s^2 =  - \D t^2 + \D x^2 + \D y^2 + \D z^2$\\ 
  &           \phantom{$ \D s^2 =  $} $  
             - 2\frac{\Phi^7}{2 \pi\;r^2}(y\; \D t + t\; \D y)\;
             (x\; \D y - y\; \D x)
             +  \left(\frac{\Phi^7}{2\pi\;r^2}\right)^2
                \left(y^2 - t^2\right)(x\; \D y - y\; \D x)^2$\\[1mm]
8 & $ \D s^2 =  - \D t^2 + \D x^2 + \D y^2 + \D z^2$\\ 
  &            \phantom{$ \D s^2 =  $} $ 
              - 2\frac{\Phi^8}{2 \pi\;r^2}(x\; \D t + t\; \D x)\;
              (x\; \D y - y\; \D x)
              +  \left(\frac{\Phi^8}{2\pi\;r^2}\right)^2
                \left(x^2 - t^2\right)(x\; \D y - y\; \D x)^2$\\[1mm]
9 & $ \D s^2 =  -\left(\D t + \frac{\Phi^9}{2\pi} z \D\phi\right)^2 + \D r^2 
              + r^2 \D\phi^2
              + \left(\D z + \frac{\Phi^9}{2\pi} t \D\phi\right)^2$\\[1mm]
\hline
\end{tabular}
\end{table}

\section{Space-\ and time-supported distortions}\label{Sec:STSupportedDis}
In order to arrive at a description of space-\ and time-supported distortions,
we must (i)\ specify the plane that will support the distortions, and (ii)\ 
consider how to adapt the Volterra process. Starting off with (i), we choose 
the $t$--$z$ plane as support. The structure of the space-\ and time-supported 
Volterra process is worked out most clearly in terms of the two new parameters
\begin{equation}\label{Eq:tauzeta} 
   \tau  \equiv  \sqrt{z^2 + t^2} \ \ {\mbox{and}}\ \ 
   \zeta  \equiv  \arctan (t/z), 
\end{equation}
which will take the place of the radial coordinate $r$ and the angular
coordinate $\phi$, respectively.\par

In analogy to what has been done to construct space-supported distortions,
we shall obtain {\em space-\ and time-supported dislocations\/} by choosing
a local translation, which is proportional to $\zeta$, the ``angle'' in the 
supporting two-plane, and {\em space-\ and time-supported 
disclinations\/} if we use a local rotation (or, more generally, a Lorentz 
boost) proportional to $\zeta$.

Presently, we shall restrict our study of space-\ and time-supported 
distortions to two cases. The remaining defects can be  
constructed in very much the same way. 
To start with, we consider the displacement field $\mcal T$ of
the time-dislocation, which must 
be adapted to the supporting $z$--$t$ plane. This is achieved by the 
displacement
\begin{equation}\label{coframe0'} 
   {\mbox{\boldmath{${\mathcal{T}}$}\unboldmath}} = 
   \frac{\chi_0}{2\pi} {\delta^i}_0 \arctan(t/z) 
   \, {\mbox{\boldmath{$e$}\unboldmath}}_i
   = \chi_0\, \zeta
   \, {\mbox{\boldmath{$e$}\unboldmath}}_0 \;, 
\end{equation} 
where $\chi_0 = \mbox{constant}$.
Having specified the appropriate gauge transformation, all geometric quantities
are computed as before. This yields the orthonormal coframe  
\begin{equation}
\left. \begin{array}{l}\displaystyle
   \omega^0 =  \D t - \frac{\chi_0}{2 \pi \tau^2} (z\;\D t - t\; \D z) 
            =  \D t - \frac{\chi_0}{2 \pi}\;\D\zeta\\
   \omega^1 = \D x \\
   \omega^2 = \D y \\
   \omega^3 = \D z \;,  
       \end{array} \right.
\end{equation}
the line element
\begin{equation} 
   \D s^2 = - \left[\D t + \frac{\chi_0}{2\pi \tau^2} 
            \left( t \D z  - z \D t \right)\right]^2 + 
            \D r^2 +  r^2 \D\phi^2 + \D z^2 \;,
\end{equation}   
and the Burgers vector
\begin{equation}
    {\mbox{\boldmath{${\mathcal{B}}$}}}=\chi_0\, {\mbox{\boldmath{$e$}}}_0 \;.
\end{equation}
It should not be surprising that the Burgers vector does not seem to depend 
on the support of the distortion, since this information has already been
used to choose the curve $S$ in (\ref{Eq:BurgersVector}) to lie in the
supporting plane, i.e.\  the $t$--$z$ plane.  
The torsion two-form, by (\ref{eq:structure1}),  has the only
nontrivial component 
\begin{equation}\label{torsiond01}
   T^0  = \D \omega^0  = -\chi_0 \,\delta^2(z,t) \, \D z \wedge \D t \;, 
\end{equation}
where we have used $\D^2 \arctan t/z = 2\pi\delta^2(z,t)\, \D z \wedge \D t$.
By evaluating the structure equation (\ref{eq:structure2}) we find that
the curvature two-form $\mbold R$ vanishes: the distorted spacetime remains
globally flat. The present space- and time-supported dislocation is 
closely related to the corresponding space-supported defect: the
torsion two-form (\ref{torsiond01}) of the former arises 
by formally replacing $(x,y)$ by $(t,z)$ in the expression (\ref{eq:torsion0})
for the torsion of latter. This correspondence clearly reflects the change
of the support from the $x$--$y$ plane to the $t$--$z$ plane.  

The space- and time-supported dislocation in the $z$-direction is
characterized by the local translation
\begin{equation} 
   {\mbox{\boldmath{${\mathcal{T}}$}\unboldmath}} = 
   \frac{\chi_3}{2\pi} {\delta^i}_3 {\;\mbox{arctan}\;}
   {\mbox{\boldmath{$e$}\unboldmath}}_i
   = \frac{\chi_3}{2\pi} \zeta
   {\mbox{\boldmath{$e$}\unboldmath}}_3 
\end{equation} 
resulting in a globally flat spacetime with the orthonormal coframe
\begin{equation}
\left. \begin{array}{l}\displaystyle
   \omega^0  =  \D t \\
   \omega^1  =  \D x \\
   \omega^2  =  \D y \\
   \omega^3  =  \D z - \frac{\chi_3}{2 \pi \tau^2}(z\; \D t - t\; \D z) 
             =  \D z - \frac{\chi_3}{2 \pi}\; \D \zeta \;,   
      \end{array} \right.
\end{equation}
which corresponds to the line element
\begin{equation} 
   \D s^2 = - \D t^2 + \D r^2 +  r^2 \D\phi^2 +   
       \left[\D z + \frac{\chi_3}{2\pi \tau^2} 
       \left(t \D z  - z \D t \right)\right]^2 \;.
\end{equation}
The Burgers vector and the torsion two-form read 
\begin{equation}
   {\mbox{\boldmath{${\mathcal{B}}$}}}=\chi_3\, {\mbox{\boldmath{$e$}}}_3 
\end{equation}
and
\begin{equation}
   T^3  = \D \omega^3  = -\chi_3 \,\delta^2(z,t) \, \D z \wedge \D t \;, 
\end{equation}
respectively. Again, the Burgers vector has the same components as in the 
case of the space-supported axial dislocation. The metrics () and (), due 
to the Lorentzian signature of the support, are nonstatic. Although the
geometry has been analyzed here, a physical interpretation must rely on
an interior solution, compare the Appendix.

\section{Field equations and matching conditions}

\subsection{Field equations}\label{sec:fieldeqs}
Formulated in the language of exterior differential calculus, the 
field equations of EC theory read \cite{Trautman73a,Baekler92}:
\begin{eqnarray}
  \frac{1}{2} \eta_{\alpha\beta\gamma} \wedge R^{\beta\gamma} 
               & = & \kappa \,\Sigma_\alpha\; , \label{eq:field1}\\
  \frac{1}{2} \eta_{\alpha\beta\gamma} \wedge T^{\gamma} 
               & = & \kappa \,\tau_{\alpha\beta}\;. \label{eq:field2}
\end{eqnarray} 
On the left-hand sides, we have the curvature two-form
$R^{\alpha\beta}$ and the torsion two-form $T^\alpha$.  The matter
currents on the right-hand sides are represented by the canonical 
energy--momentum three-form $\Sigma_\alpha$ and the spin
three-form $\tau_{\alpha\beta}$. The Einsteinian
gravitational constant is denoted by $\kappa$. 
The field equations (\ref{eq:field1}) and (\ref{eq:field2}) may be
cast into an `effective' Einsteinian form \cite{Hehl76}. It has been
argued \cite{Hehl74,Kerlick75} that the corresponding `effective'
energy--momentum current should replace the symmetric energy--momentum current
of GR when questions such as energy conditions are investigated. More recent 
investigations \cite{Smalley94} come to a similar conclusion.\par

In a static situation, the most natural procedure to find (physically
sensible) solutions to the field equations (\ref{eq:field1}) and 
(\ref{eq:field2}) of EC theory is to first specify the matter currents 
$\Sigma_\alpha$ and $\tau_{\alpha\beta}$ which are then put into
the field equations, thereby determining the geometry of spacetime,
i.e.\ the quantities $R^{\alpha\beta}$ and $T^\alpha$. For obvious
reasons, this is not a viable procedure for our present investigation.
Nevertheless, a matter model is indispensable for a physical interpretation
of the spacetimes considered. As the most prominent example
of such a matter model, we define matter currents of a {\em spin fluid\/}
by using the quite general hyperfluid paradigm by Obukhov and 
Tresguerres \cite{Obukhov93} (see also \cite{Obukhov96}), which specializes to 
\begin{eqnarray}
   \Sigma_\alpha      & = & \epsilon\, u_\alpha\, u + p(\eta_\alpha 
                            +  u_\alpha\,u)
                            - 2  u^\beta \dot{S_\alpha}_\beta\,u\; ,
                            \label{eq:matcur1}\\
   \tau^{\alpha\beta} & = & S^{\alpha\beta} u \; \label{eq:matcur2}
\end{eqnarray}
for the matter currents of a spin fluid. The flow three-form $u$ must be 
normalized, ${\,\!}^* u \wedge u = - \eta$, and is related to the usual 
velocity four-vector $u=u^\alpha\,e_\alpha$ by
\begin{equation}\label{eq:ual}
   u^\alpha = e^\alpha \inner ({\,\!}^* u) \;.
\end{equation}
The rest-energy density $\epsilon$, the spin density ${S_\alpha}_\beta$, and
the particle density $\rho$ are related by an equation of state 
$\epsilon = \epsilon(\rho, {S_\alpha}_\beta)$ that must be chosen 
appropriately.  

\subsection{Matching conditions}\label{sec:matching}
The AKP matching conditions of the EC theory determine under which conditions 
a given matter distribution
may be matched consistently---i.e.\ such that the field equations 
(\ref{eq:field1}) and (\ref{eq:field2}) are fulfilled in the distributional 
sense---to a vacuum solution. \par

The original formulation of Arkuszewski, Kopczy{\'n}ski and Ponomariev
\cite{Arkuszewski75}, see also Chmie\-lowski \cite{Chmielowski92},
takes recourse to the coordinate components of the geometrical and
dynamical objects involved. In other words, everything is referred to a
{\it natural} frame $\partial_i$ or coframe $\D x^j$, respectively. Then,
instead of the coframe, the components of the metric
$g_{ij}=e_i{}^\alpha e_j{}^\beta o_{\alpha\beta}$, together with those
of the contortion $K_{kij}=e_i{}^\alpha e_j{}^\beta
K_{k\alpha\beta}\,,$ enter the scene. Torsion and contortion are
interrelated according to $T^\alpha=K^\alpha{}_\beta\wedge\omega^\beta$ with
$K^\alpha{}_\beta=K_k{}^\alpha{}_{\beta}\,\D x^k\,.$ The connection
can be written as the
sum of the Christoffel symbol and a contortion
part:
\begin{equation}
   \Gamma_{ij}{}^k = \Gamma^{\{\}}_{ij}{}^k-K_{ij}{}^k\;.
\end{equation}
The energy--momentum and spin three-forms, in components with respect to
the $\eta$-basis, read
\begin{equation}
  \Sigma_\alpha      = \Sigma_\alpha{}^k\,\eta_k\;,\qquad
  \tau_{\alpha\beta} = \tau_{\alpha\beta}{}^k\,\eta_k \;,
\end{equation}
where $\Sigma_\alpha{}^k$ and $\tau_{\alpha\beta}{}^k$ are the canonical 
tensors of energy-momentum and spin, respectively.\par

Let us suppose that there exists a coordinate $x$ defining locally a
hypersurface $\cal S$ by the equation $x = 0$, such that matter is
bound to the region $x < 0$.  Furthermore, we define the normalized
vector field with components $n_i = \partial_i x / | \nabla x |$ and the
operator 
\begin{equation}
   {h^i}_j = \delta^i_j + n^i n_j\; , 
\end{equation}
projecting on
hypersurfaces $x = \mbox{const}$. For any geometric object $T$, we may use
${h^i}_j$ to construct its projection on the hypersurface $x =
\mbox{const}$, which will be denoted by $\overline T$. We will also use the
notation $T|_{\pm} \equiv \lim_{x \rightarrow \pm 0} T$.\par

For a matter distribution, the components ${\Sigma}_{i}{}^k$ and
$\tau_{ij}{}^k$ of the energy--momentum tensor and the spin tensor
are subject to the {\em junction conditions\/} \cite{Arkuszewski75}
\begin{equation}\label{eq:AKPcond1}
  \left. n_k \,\Sigma_{i}{}^k \right|_{-} - \left. n_i\, {\overline K}_{jkl}
    {\overline K}^{klj} \right|_{-} = 0 
\end{equation} 
and 
\begin{equation}\label{eq:AKPcond2}
  \left. n_k\, \tau_{ij}{}^k \right|_{-} = 0 \; , 
\end{equation}
respectively. If the conditions (\ref{eq:AKPcond1}) and (\ref{eq:AKPcond2}) 
are fulfilled, then there exists a unique solution to the vacuum field
equations, which is determined by the {\em boundary conditions\/} 
\begin{equation}\label{eq:AKPbound1}
  \left. g_{ij} \right|_{+} = \left. g_{ij} \right|_{-} 
\end{equation}
and 
\begin{equation}\label{eq:AKPbound2}
  \left. \partial_x g_{ij} \right|_{+} = \left. \partial_x g_{ij} \right|_{-} +
  \left. 2 {K}_{(ij)}{}^x/g^{xx} \right|_{-}\; , 
\end{equation}
restricting the
components and the first derivatives of the metric tensor $g_{ij}$.
Therefore the metric tensor is continuous across the bounding
hypersurface, whereas its first derivatives with respect to the
function $x$ have a jump that
is determined by the contortion
components ${K}_{ij}{}^k$.  This last fact clearly distinguishes the
AKP matching conditions from the Lichnerowicz matching conditions of GR.
\section{Extended matter sources}
\label{Sect:Extended}

By generalizing the Volterra process to four-dimensional spacetime we 
have generated all possible vacuum geometries outside an infinitely long
straight-line defect. In this section we shall find extended matter 
distributions producing the same exterior metrics as the Volterra processes
zero, three, and six.
The canonical energy--momentum tensor is assumed to take a string-like form
satisfying the weak energy condition by construction.
However, the metric energy--momentum tensor may be subject to
stronger restrictions which put a limit on the strength of torsion \cite{Soleng:PRD94}.
 
\subsection{Cosmic string analogue to a spinning particle in 2+1 dimensions}
\label{Sect:Spinning particle}

A point particle in (2+1)-dimensional gravity translates into an 
infinitely thin, straight cosmic string in (3+1)-dimensional spacetime. 
If the point particle is permitted to have spin, then the lift from 
2+1 to 3+1 dimensions is nonunique, since the spin-tensor component 
is undetermined along the additional spatial dimension. 
Thus, there are three physically distinct types
of cosmic strings with spin \cite{Gal'tsov93}. 

Here we shall derive an interior solution, which corresponds to 
these three types of spinning line-like topological defects:
the spinning cosmic string \cite{Mazur86}, 
the cosmic dislocation
(twisted string) \cite{Gal'tsov93},
and the cosmic string interacting with a gravitational wave. 
The interior spinning string solution \cite{Soleng92a}
is a special case of this solution. 

Matching of the solution to an exterior Einstein vacuum leads to the general
exterior solution that was studied in detail in Ref.~\cite{Gal'tsov93}.

\subsubsection{Geometry} 

Let the line element of spacetime be of the form 
\begin{equation}
   \D s^2 = 
           - [\D t + M(r) \D\phi]^2 
           + \D r^2 + \rho(r)^2 \D\phi^2 + [\D z + C(r) \D\phi]^2 \;. 
\label{Eq:MetricAnsatz}
\end{equation}  
If $\rho(r)=r$ and $C(r)=M(r)=0$, Eq.~(\ref{Eq:MetricAnsatz}) reduces
to the line element of Minkowski spacetime in terms of cylindrical coordinates.

Let us define a (pseudo-)orthonormal 
frame specified by the following one-form basis:
\begin{equation}
   \left.
       \begin{array}{l}\displaystyle  
       \omega^{\hat 0} = \D t + M(r)\D \phi \\ 
       \omega^{\hat 1} = \D r\\ 
       \omega^{\hat 2} = \rho(r)\D\phi\\
       \omega^{\hat 3} = \D z + C(r)\D \phi \;.\\ 
       \end{array}
   \right.
   \label{Eq:BasisOneForms}
\end{equation}
Let the nonvanishing components of the energy-momentum current  
be given as
\begin{equation}\label{Eq:EnergyMomentum}
    \Sigma^{\hat 0} = -\lambda\, \omega^{\hat 1}\wedge \omega^{\hat 2}
                                 \wedge\omega^{\hat 3}\;,\quad
    \Sigma^{\hat 3} = -\lambda\, \omega^{\hat 0}\wedge \omega^{\hat 1}
                                 \wedge\omega^{\hat 2}\;,
\end{equation}
where $\lambda$ is a constant, in accordance with the homogeneity 
assumption. The spin three-form is assumed to have the only nonzero 
component 
\begin{eqnarray}
  \tau^{\hat 1 \hat 2} \;=\; -\tau^{\hat 2 \hat 1} 
                       & = &  \sigma\, \eta^{\hat 0} + \beta\, \eta^{\hat 3} 
                              \nonumber\\
                       & = & -\sigma\, \omega^{\hat 1}\wedge \omega^{\hat 2}
                                \wedge\omega^{\hat 3}
                             -\beta\, \omega^{\hat 0}\wedge \omega^{\hat 1}
                                \wedge\omega^{\hat 2} \;.\label{Eq:Spin}
\end{eqnarray} 
Also $\beta$ and $\sigma$ are constants.
Note that unlike the energy--momentum current,
the components of the spin current are {\em not\/}
boost-invariant along the axis of symmetry. Hence,
if we define another Lorentz frame by
a boost along the cylinder
\[
    \left(\begin{array}{c} 
       \omega^{\hat 0'}\\  
       \omega^{\hat 1'}\\  
       \omega^{\hat 2'}\\  
       \omega^{\hat 3'}  
          \end{array}
    \right) 
           = \left(
    \begin{array}{c c c c}
     \gamma(v) & 0 & 0 & -v\,\gamma(v) \\
     0 & 1 & 0 & 0  \\ 
     0 & 0 & 1 & 0  \\ 
    -v\,\gamma(v) & 0 & 0 & \gamma(v)   
    \end{array} 
    \right) 
    \left(\begin{array}{c} 
       \omega^{\hat 0}\\  
       \omega^{\hat 1}\\  
       \omega^{\hat 2}\\  
       \omega^{\hat 3}  
          \end{array}
    \right)\;, 
\]
then the spin three-form is given by
\begin{eqnarray*}
  \tau^{\hat 1' \hat 2'} & = & -\sigma'\, \omega^{\hat 1'}\wedge 
                                \omega^{\hat 2'}\wedge\omega^{\hat 3'}
                             -\beta'\, \omega^{\hat 0'}\wedge 
                                \omega^{\hat 1'}\wedge\omega^{\hat 2'}\\
                         & = & -\gamma(v)(\sigma-v\beta)\,\omega^{\hat 1'}
                                \wedge \omega^{\hat 2'}\wedge\omega^{\hat 3'}
                               -\gamma(v)(\beta - v\sigma)\, \omega^{\hat 0'}
                               \wedge\omega^{\hat 1'}\wedge\omega^{\hat 2'}\;.
\end{eqnarray*} 
{}From these relations it is clear that
$\beta^2-\sigma^2$ is an invariant under boosts 
along the cylinder. This quantity is interpreted quite naturally within the 
spin-fluid model of Sect.~\ref{sec:fieldeqs}.
The spin current (\ref{eq:matcur2}) specializes to our ansatz (\ref{Eq:Spin}) 
if the flow three-form and the spin density are given by  
$u = \gamma(v)(\eta^{\hat 0} + v\,\eta^{\hat 3})$ and $S^{\alpha\beta} = 
s\, \delta^{[ \alpha}_{\hat 1} \delta^{\beta]}_{\hat 2}$, respectively.
Therefore 
\begin{equation}\label{sigbeta}
   \sigma = -\gamma(v) \, s\;,\quad \beta = \gamma(v) v \,s\;,
\end{equation}
and the given invariant turns out to be the spin density $s$
in the rest frame of the fluid, 
\begin{equation} 
  s = \sqrt{\beta^2-\sigma^2}\;.
\end{equation}

\subsubsection{Interior solution}\label{s:interior}

For computational simplicity we define the zero-forms
\begin{equation}\label{eq:defomups}
    \Omega(r) \equiv  \kappa\,\sigma + \frac{M'}{2 \rho} 
    \quad\quad\mbox{and}\quad\quad
    \Upsilon(r) \equiv \kappa\,\beta -  \frac{C'}{2 \rho}\;.
\end{equation}
The prime denotes a partial derivative with respect to $r$.
According to the second field equation (\ref{eq:field2}), 
the torsion two-form for the source (\ref{Eq:Spin}) is given by
\begin{equation}
  \begin{array}{l}
   T^{\hat 0}  =  2\kappa\sigma\, \eta^{\hat 0 \hat 3} 
               =   - 2\kappa\sigma\, \omega^{\hat 1} \wedge\omega^{\hat 2}\;,\\
   T^{\hat 3}  =  -2\kappa\beta\, \eta^{\hat 0 \hat 3} 
               =   2\kappa\beta\, \omega^{\hat 1} \wedge\omega^{\hat 2}\;.
   \end{array}
\label{Eq:Torsion}
\end{equation} 
The Einstein three-form computed from the metric (\ref{Eq:MetricAnsatz})\ 
and the torsion (\ref{Eq:Torsion})\ reads
\begin{eqnarray*}
   G^{\hat 0} &=& \left(\frac{1}{\rho}\left(C'\Upsilon + M'\Omega-\rho''\right)
                        + \Omega^2 + \Upsilon^2 \right) 
               \omega^{\hat 1}\wedge\omega^{\hat 2} \wedge\omega^{\hat 3}
              + 2 \Omega\Upsilon\, 
               \omega^{\hat 0}\wedge\omega^{\hat 1} \wedge\omega^{\hat 2}
              - \Omega'\,
               \omega^{\hat 0}\wedge\omega^{\hat 1} \wedge\omega^{\hat 3}\\
   G^{\hat 1} &=& \left( -\Omega^2 + \Upsilon^2\right)
               \omega^{\hat 0}\wedge\omega^{\hat 2} \wedge\omega^{\hat 3}\\
   G^{\hat 2} &=& - \Omega'\,
               \omega^{\hat 1}\wedge\omega^{\hat 2} \wedge\omega^{\hat 3}
                  - \Upsilon'\,
               \omega^{\hat 0}\wedge\omega^{\hat 1} \wedge\omega^{\hat 2}
               + \left(\Omega^2 - \Upsilon^2\right)
               \omega^{\hat 0}\wedge\omega^{\hat 1} \wedge\omega^{\hat 3}\\
   G^{\hat 3} &=& - 2 \Omega\Upsilon \,
               \omega^{\hat 1}\wedge\omega^{\hat 2} \wedge\omega^{\hat 3}
                + \left(\frac{1}{\rho}\left(C'\Upsilon + M'\Omega-\rho''\right)
                        - \Omega^2 - \Upsilon^2 \right) 
               \omega^{\hat 0}\wedge\omega^{\hat 1} \wedge\omega^{\hat 2}
              + \Upsilon'\,
               \omega^{\hat 0}\wedge\omega^{\hat 1} \wedge\omega^{\hat 3}\;.
\end{eqnarray*}
Considering the energy--momentum current as defined in 
(\ref{Eq:EnergyMomentum}), the first field equation (\ref{eq:field1}) is 
solved by
\begin{equation}
    \Upsilon(r) = \Omega(r) = 0
    \label{Eq:G0}
\end{equation}
and 
\begin{equation}
    \rho = \frac{1}{\sqrt{\lambda}} \sin (\sqrt{\lambda} r)\;,
    \label{Eq:SolutionRho}
\end{equation}
where a constant of integration has been determined by the requirement 
that the metric must be Minkowskian in the limit $r \rightarrow 0$.
Using (\ref{Eq:G0})\ and (\ref{Eq:SolutionRho})\ 
with the boundary conditions $M(0)=C(0)=0$, the expressions (\ref{eq:defomups})
can be integrated to give the explicit forms of the functions 
$C(r)$ and $M(r)$, respectively: $M(r)=\sigma \int_0^r \rho\, \D r'$ and 
$C(r)=\beta\int_0^r \rho\,\D r'$. 
Using (\ref{Eq:SolutionRho}), these functions are given implicitly by
\begin{equation}
    \lambda \int_0^r \rho(r')\, \D r'= 1-\cos(\sqrt{\lambda} r) \;. 
\end{equation} 
The expressions for $M(r)$ and $C(r)$ should be compared with
the string's mass. Its mass per unit length is found by
integrating the density over a cross section of the cylinder, 
\begin{equation}
  \mu\equiv \lambda \int_{0}^{2\pi}\int_{0}^{R}
            \omega^{\hat 1}\wedge\omega^{\hat 2}  
     = \lambda\int_{0}^{2\pi}\int_{0}^{R}\rho(r)\, \D r\wedge \D\phi \;, 
  \label{Eq:muintegral} 
\end{equation}
where $R$ is the radial coordinate distance from the origin to the 
surface of the cylinder. Likewise, the two components of the spin 
are given by similar integrals with spin densities replacing the 
energy density:
\begin{mathletters}
\begin{eqnarray}
   j^{\hat 0} &\equiv& \sigma\int_{0}^{2\pi}\int_{0}^{R}\rho(r)\, 
   \D r\wedge  \D\phi \;,\label{Eq:j0}\\ 
   j^{\hat 3} &\equiv& \beta\int_{0}^{2\pi}\int_{0}^{R}\rho(r)\, 
   \D r\wedge  \D\phi \;.\label{Eq:j3} 
\end{eqnarray}
\end{mathletters}%
Note that these two equations can be expressed more compactly by means
of the surface values of the metric coefficients $M(r)$ and $C(r)$. 
Hence, the results are
\begin{mathletters}
\begin{eqnarray}
   \mu   &=& 2\pi \left[1-\cos(\sqrt{\lambda} R)\right] \;,\\
   j^{\hat 0} &=& 2\pi M(R) \;,\\
   j^{\hat 3} &=& 2\pi C(R) \;,
\end{eqnarray}
\end{mathletters}%
\setcounter{Results}{\value{equation}}%
where $j^{\mu}$ are the components of a vector. These relations make up 
the boundary conditions for the exterior gravitational field that we will
construct in the next section. 

With the field equations solved, the curvature two-form is left with
the only nonvanishing components 
\begin{equation}\label{eq:curvInterior}
   R^{\hat 1}{}_{\hat 2} = -R^{\hat 2}{}_{\hat 1} 
   = \lambda \,\omega^{\hat 1} \wedge \omega^{\hat 2}\;,
\end{equation}
which is the finite version of expression (\ref{eq:curvOrder6}) 
for the wedge disclination (order six).  

\subsubsection{Exterior solution}\label{sec:exterior}

We shall now use the AKP conditions as formulated in  Sect.~\ref{sec:matching}
and find the exterior solution corresponding to the above spin-polarized 
cylinder.  By construction, the vacuum solution will comprise the two screw 
dislocations (orders zero and three) and the wedge disclination (order six). 
Hence, we use the line element
\begin{equation}\label{Eq:ExteriorMetric}
   g =  - \left( \D t + \frac{\Theta^0}{2\pi}\,\D\phi \right)^2 + \D r^2
        + \left(1 - \frac{\Phi^6}{2\pi}\right)^2 (r+r_{0})^2\D\phi^2 
        + \left( \D z + \frac{\Theta^3}{2\pi}\,\D\phi \right)^2\,,
\end{equation}
where the constant parameters $\Theta^0$, $\Theta^3$, and $\Phi^6$ have the 
same  interpretation as in Sect.~\ref{s:volterra} and belong to the three 
distortions under consideration. The spacetime with the (locally) flat metric 
(\ref{Eq:ExteriorMetric}) has also been investigated recently by 
Tod \cite{Tod94} 
and by Gal'tsov and Letelier \cite{Gal'tsov93}. By straightforward 
computation we find that (\ref{eq:AKPcond1}) and (\ref{eq:AKPcond2}) are 
satisfied for the present matter currents (\ref{Eq:EnergyMomentum}) 
and (\ref{Eq:Spin}). The junction conditions (\ref{eq:AKPbound1}) 
and (\ref{eq:AKPbound2}) read 
\begin{eqnarray*}
   \left.g_{\phi\phi}\right|_{{}_{+}}&=&\left.g_{\phi\phi}\right|_{{}_{-}}\;,\\
   \left.g_{\phi z}\right|_{{}_{+}}&=&\left.g_{\phi z}\right|_{{}_{-}}\;,\\
   \left.g_{\phi t}\right|_{+}&=&\left.g_{\phi t}\right|_{-}\;,\nonumber\\
   \left.g_{\phi\phi,r}\right|_{{}_{+}}&=&
   \left.g_{\phi\phi,r}\right|_{{}_{-}}-2K_{\phi\phi}{}^r\;,\\
   \left.g_{\phi z,r}\right|_{{}_{+}}&=&
   \left.g_{\phi z,r}\right|_{{}_{-}}-2K_{(\phi z)}{}^r\;,\\
   \left.g_{\phi t,r}\right|_{{}_{+}}&=&
   \left.g_{\phi t,r}\right|_{{}_{-}}-2K_{(\phi t)}{}^r\; . 
\end{eqnarray*} 
Written out explicitly, we have the following system of algebraic
equations for determining $a$, $B$, $r_0$, and $b$:
\begin{eqnarray*}
   \left(1 - \frac{\Phi^6}{2\pi}\right)^2 (R+r_{0})^2+b^2-a^2
      &=&\rho(R)^2+C(R)^2-M(R)^2\;,\\
   \frac{\Theta^0}{2\pi} &=& M(R)\;,\\  
   \frac{\Theta^3}{2\pi} &=& C(R)\;,\\
   2 \left(1 - \frac{\Phi^6}{2\pi}\right)^2 (R+r_{0})
      &=& 2\rho'(R)\rho(R)+2C'(R)C(R) -2 M'(R) M(R) \nonumber\\
      & & {} + 2\rho (R) [\sigma M(R)-\beta C(R)]\;,\\
   0 &=& C'(R)-\beta \rho(R)\;,\\
   0 &=&-M'(R)+\sigma \rho(R)\;.
\end{eqnarray*}
Combining the above conditions with (\ref{Eq:SolutionRho})\ and 
(\theResults)\  we get
\begin{eqnarray}
   \left(1-\frac{\Phi^6}{2\pi}\right)&=&\cos(\sqrt{\lambda}R)=1-\mu/(2\pi)\;,
   \label{sol1}\\
   \frac{\Theta^0}{2\pi}&=& M(R)=\frac{j^{\hat 0}}{2\pi}=-\frac{2 \gamma(v)s}
   {\lambda}\left(\cos(\sqrt{\lambda}\,R)-1\right)\;, \label{sol2}\\
   \frac{\Theta^3}{2\pi}&=& C(R)=\frac{j^{\hat 3}}{2\pi}=-\frac{2 \gamma(v)vs}
   {\lambda}\left(\cos(\sqrt{\lambda}\,R)-1\right)\;,\label{sol3}\\
   r_{0}&=& \frac{1}{\sqrt{\lambda}}\tan(\sqrt{\lambda} R)-R\;,
\end{eqnarray}
where we have evaluated the integrals (\ref{Eq:j0}) and (\ref{Eq:j3}) 
with (\ref{sigbeta}) inserted.
The solution falls into three classes depending on the nature of the 
vector $j^{\mu}$. If it is timelike, the solution corresponds to the
spinning string \cite{Mazur86,Soleng92a}. If it is null, the metric can
be regarded as the limiting case of a cosmic string interacting with
a gravitational wave, and if it is spacelike, the solution is a so-called
cosmic dislocation \cite{Gal'tsov93}. 

The mass and spin 
parameters of the source can be related to dislocations and 
disclinations as described in Section~\ref{Sect:Distortions}.
Hence, the mass per length $\mu$ produces an axial disclination  
(distortion of order six) with disclination strength $\Phi^6=\mu$, see
Table \ref{tab:disclin}; the zeroth component of the spin vector 
$j^{\hat 0}$ induces a time-dislocation (order zero) and $j^{\hat 3}$ 
results in an axial dislocation (order three); the corresponding
exterior metrics are those given in Table \ref{tab:disloc}.
The exterior metric (\ref{Eq:ExteriorMetric})\ can thus be produced by  
a combination of these spacetime distortions. 

The deficit angle $\Delta\phi$ outside the string is determined by
$\Delta\phi=2\pi(1- B)=\mu$. Note that in the conical case
there is a maximal radius of the cylinder given by
\begin{equation}
    R=\frac{\pi}{2} \lambda^{-1/2}\;.  
\end{equation}
This maximal radius corresponds 
to a string with deficit angle and mass per length equal to $2\pi$.
In this case, $r_{0}$ diverges, and in the first term of $g_{\phi\phi}$ 
in the exterior metric, only the combination $B^2 r_{0}^2$ survives, 
making $g_{\phi\phi}$ constant here. Thus, for $\mu =2\pi$, spacetime 
becomes cylindrical rather than conical.

\section{Concluding remarks}

A Poincar\'e gauge theory is a natural four-dimensional generalization 
of the $SO(3) {\;{\rlap{$\supset$}\times}\;} T(3)$ gauge theory of 
solid continua with line defects. {}From this perspective we have found 
spacetime equivalents of the Volterra distortions of elastic media.
For a subset of these spacetime distortions we have found the corresponding 
nonsingular sources in the Einstein--Cartan theory of gravity.

If we recall the interpretation of the interior solution in terms
of the convective spin fluid under consideration---see Sect.~\ref{sec:fieldeqs}
and the remarks at the end of Sect.~\ref{s:interior}---we find that 
the geometry of the combined interior and exterior spacetime can be 
formulated with respect to two complementary sets of parameters. Starting 
with the interior spacetime, the relevant set is 
$(\lambda,s,v) = (\text{mass density},\text{spin density},\text{velocity})$;
we call this the {\em physical} set. If, on the other hand, we start with 
the vacuum geometry, we should use the distortion strengths 
$(\Theta^0,\Theta^3,\Phi^6)$; this is the {\em geometrical} set of parameters.
While expressions (\ref{sol1})--(\ref{sol3}) represent the geometrical set 
in terms of the physical set, the complementary relations are
\begin{align}
   \lambda &= \frac{1}{R^2}\acos^2\left(1-\frac{\Phi^6}{2\pi} \right)\,,
   \label{isol1}\\
    s &= \frac{\sqrt{(\Theta^0)^2 -(\Theta^3)^2}}{2R^2(\Phi^6 -4\pi)}\,
         \acos\left(1-\frac{\Phi^6}{2\pi}\right)\,,\label{isol2}\\
    v &= -\frac{\Theta^3}{\Theta^0}\,.\label{isol3}
\end{align}
Correspondingly, all relevant geometrical quantities can be formulated 
with respect to either $(\lambda,s,v)$ or $(\Theta^0,\Theta^3,\Phi^6)$.
In particular, the time dislocation and the axial dislocation are no 
longer independent of each other: by (\ref{isol3}), the corresponding 
dislocation strengths are subject to the relation $\Theta^3 = -v\,\Theta^0$.
For instance, if $\Theta^3$ is fixed to some finite value, then 
$\Theta^0\rightarrow 0$ would imply $v\rightarrow\infty$, which is
clearly forbidden on physical grounds, since the fluid velocity must not 
exceed the speed of light. Thereby we conclude that the axial dislocation 
cannot occur alone, it is always accompanied by a time dislocation. 
Moreover, by the same token, the `cosmic dislocation' \cite{Gal'tsov93}, 
is unphysical in the sense that it corresponds to a spacelike 
four-velocity of the spin fluid.

Whether interior solutions can be found for the remaining (space- or 
space- and time-supported) distorted vacua remains an open question 
and deserves further investigation. The present strategy for constructing
string sources is, at least in principle, applicable to these geometries, too:
(i) take the {\em coframe} and replace the constant distortion strength
by a {\em function}, (ii) find the nonvanishing components of
the {\em torsion\/} two-form and make a corresponding ansatz for the 
interior spacetime, and (iii) solve the field equations. {\em If\/}
a solution is found, then, by construction, it is expected to reproduce
the respective distorted vacuum. It turns out, however, that in the generic
case we have to face two problems that didn't appear in the cases solved.

First of all, most of the remaining
distorted vacua are either not cylindrically symmetric or nonstatic,
compare Tables \ref{tab:disloc} and \ref{tab:disclin}. Secondly, 
the structure of the {\em second\/} field equation, with the 
the present spin fluid model, only allows for a quite 
restricted torsion part---as demonstrated in the Appendix for one
of the space- and time-supported defects---of the interior geometry.
While the first mentioned problem merely is of technical nature,
the circumvention of the second one requires substantial changes 
of the physical models used. One possible modification could be
to consider a different spin-fluid matter model, such as that 
developed by Ray and Smalley \cite{Ray82}, which was recently used
by Krisch \cite{Krisch96} to find a family of spinning string-type 
solutions in a spacetime with curvature and torsion.

\subsection*{Acknowledgements}
We thank F.W. Hehl for critically reading the manuscript and F. Gronwald
and Yu.N. Obukhov for interesting discussions.   
Most of the computations have been checked with the aid of 
REDUCE~\cite{Hearn93} making use of the Excalc package~\cite{Schrufer94}.
The results of Sect.~\ref{Sect:Spinning particle}\ 
have been found using the tensor algebra package {\sc Cartan}~\cite{Soleng94}.
HHS thanks F.W. Hehl and the {\em Graduiertenkolleg Scientific Computing},
University of Cologne, for hospitality during the initial stage of this work.
One of the authors (RAP) is supported by the {\em Graduiertenkolleg
Scientific Computing}, 
University of Cologne and GMD---German National 
Research Center for Information Technology, St. Augustin.
\section*{Appendix}
In Sect.~\ref{Sect:Extended} we have demonstrated how to find, in the
framework of EC theory, extended matter sources corresponding to 
the vacuum geometries of space-supported distorted spacetimes. Applying
the same scheme to the space- and time-supported vacuum defects of
Sect.~\ref{Sec:STSupportedDis}, we proceed as follows. Define the 
orthonormal coframe 
\begin{equation}\label{coframe01}
   \left.
       \begin{array}{l}\displaystyle  
       \omega^{\hat 0} =  \D t - \frac{E}{2 \pi (t^2+z^2)} (z\;\D t - t\; \D z) 
                       =  \D t - \frac{E}{2 \pi}\;\D\zeta\\
       \omega^{\hat 1} = \D r\\ 
       \omega^{\hat 2} = r \D\phi\\
       \omega^{\hat 3} = \D z\;,\\ 
       \end{array}
   \right.
\end{equation}
which can be obtained from (\ref{coframe0'}) by replacing the constant
dislocation strength $\chi_0$ by the function $E=E(t,z)$. 
In view of the (fixed) exterior geometry that was determined in 
Sect.~\ref{Sec:STSupportedDis}, the torsion two-form is required 
to have the only nontrivial component
\begin{equation}\label{Eq:torsion01}
   T^{\hat 0} = 2\kappa\alpha \,\eta^{\hat 1\hat 2} 
   = 2\kappa\alpha \,\omega^{\hat 0}\wedge\omega^{\hat 3}\;,
\end{equation}
with some function $\alpha=\alpha(t,z)$. The resulting curvature two-form
has the nontrivial components 
\begin{equation}
   R^{\hat 0}{}_{\hat 3} = R^{\hat 3}{}_{\hat 0} = 
   \frac{\Xi\,(t E_t + z E_z) + E\,(t \Xi_t + z \Xi_z) + \Xi_z(t^2+z^2)}
   {z E +  t^2+z^2}\, \omega^{\hat 0}\wedge\omega^{\hat 3} \;,
\end{equation}
where we have introduced the function
\begin{equation}
   \Xi \equiv 2\kappa\alpha + \frac{t E_t+z E_z}{z E+ t^2+z^2 }\;.
\end{equation}
With ansatz (\ref{Eq:torsion01}), the second field equation, after some 
algebra, implies 
\begin{equation}\label{canspin}
   \tau^{\hat 1\hat3} = -\kappa\alpha\,\eta^{\hat1}\;,\quad
   \tau^{\hat 2\hat3} = -\kappa\alpha\,\eta^{\hat2}\;,
\end{equation}
while all other components of the canonical spin three-form vanish.
The form (\ref{canspin}) of the spin current, contrary to the situation
we had in Sect.~\ref{Sect:Extended}, is out of the scope of the present
matter model, since (\ref{eq:matcur2}) would imply $S^{\hat 1\hat3} = 
S^{\hat 2\hat3} = -\kappa\alpha$ for the spin density, which clearly is 
possible, and both $u = \eta^{\hat 1}$ and $u = \eta^{\hat 2}$ for the
flow three-form $u$, which is 
impossible since generally $\eta^\alpha\neq\eta^\beta$ for $\alpha\neq\beta$.
A similar argument holds for the space-supported dislocations of order 
one and two. Let us reiterate, though, that this inconvenience is due 
to the specific matter model used here, and it remains to be asked if
a differently defined spin fluid behaves differently in this respect.



\small
\begin{thebibliography}{10}

\bibitem{Vilenkin81}
A.~Vilenkin.
\newblock Gravitational field of vacuum domain walls and strings.
\newblock {\em Phys. Rev. D}, 23:852--857, 1981.

\bibitem{Gibbons90}
G.W. Gibbons, S.W. Hawking, and T.~Vachaspati, editors.
\newblock {\em The Formation and Evolution of Cosmic Strings}.
\newblock Cambridge University Press, Cambridge, 1990.

\bibitem{Vilenkin94}
A.~Vilenkin and E.P.S. Shellard.
\newblock {\em Cosmic Strings and Other Topological Defects}.
\newblock Cambridge University Press, Cambridge, 1994.

\bibitem{Hiscock85}
W.A. Hiscock.
\newblock Exact gravitational field of a string.
\newblock {\em Phys. Rev. D}, 31:3288--3290, 1985.

\bibitem{Vickers87}
J.A.G. Vickers.
\newblock Generalized cosmic strings.
\newblock {\em Class. Quantum Grav.}, 4:1--9, 1987.

\bibitem{Hehl76}
F.W. Hehl, P.~von~der Heyde, G.D. Kerlick, and J.~Nester.
\newblock General relativity with spin and torsion: Foundations and prospects.
\newblock {\em Rev. Mod. Phys.}, 48:393--416, 1976.

\bibitem{Bekenstein92}
J.D. Bekenstein.
\newblock Chiral cosmic strings.
\newblock {\em Phys. Rev. D}, 45:2794--2801, 1992.

\bibitem{Letelier95a}
P.S. Letelier.
\newblock Spinning cosmic strings as torsion line spacetime defects.
\newblock {\em Class. Quantum Grav.}, 12:471--478, 1995.

\bibitem{Holz92}
A.~Holz.
\newblock Topological properties of linked disclinations and dislocations in
  solid continua.
\newblock {\em J. Phys. A: Math. Gen.}, 25:L1--L10, 1992.

\bibitem{Gal'tsov93}
D.V. Gal'tsov and P.S. Letelier.
\newblock Spinning strings and cosmic dislocations.
\newblock {\em Phys. Rev. D}, 47:2473--4276, 1993.

\bibitem{Edelen94}
D.G.B. Edelen.
\newblock Space--time defect solutions of the {E}instein field equations.
\newblock {\em Int. J. Theor. Phys.}, 33:1315--1334, 1994.

\bibitem{Tod94}
K.P. Tod.
\newblock Conical singularities and torsion.
\newblock {\em Class. Quantum Grav.}, 11:1331--1339, 1994.

\bibitem{Holz88}
A.~Holz.
\newblock Geometry and action of arrays of disclinations in crystals and
  relation to (2+1)-dimensional gravitation.
\newblock {\em Class. Quantum Grav.}, 5:1259--1282, 1988.

\bibitem{Gerbert90}
P.~de~Sousa~Gerbert.
\newblock On spin and (quantum) gravity in (2 + 1) dimensions.
\newblock {\em Nucl. Phys. B}, 346:440--472, 1990.

\bibitem{Kohler95}
C.~Kohler.
\newblock Point particles in $2 + 1$ dimensional gravity as defects in solid
  continua.
\newblock {\em Class. Quantum Grav.}, 12:L11--L15, 1995.

\bibitem{Kohler95b}
C.~Kohler.
\newblock Line defects in solid continua and point particles in $(2 +
  1)$-dimensional gravity.
\newblock {\em Class. Quantum Grav.}, 12:2977--2993, 1995.

\bibitem{Volterra07}
V.~Volterra.
\newblock Sur l'\'equilibre des corps \'elastiques multiplement connexes.
\newblock {\em Ann. \'Ec. Norm. Sup.}, 24:401--517, 1907.

\bibitem{Nabarro67}
F.R.N. Nabarro.
\newblock {\em Theory of Crystal Dislocations}.
\newblock The International Series of Monographs on Physics. Oxford University
  Press, Oxford, 1967.
\newblock Section 8.4.3 on timelike dislocations.

\bibitem{Kleman80}
M.~Kl{\'e}man.
\newblock The general theory of disclinations.
\newblock In F.R.N. Nabarro, editor, {\em Dislocations in Solids}, volume~5,
  pages 243--297. North-Holland, Amsterdam, 1980.

\bibitem{Kroner81}
E.~Kr{\"o}ner.
\newblock Continuum theory of defects.
\newblock In R.~Balian, M.~Kl{\'e}man, and J.-P. Poirier, editors, {\em Physics
  of Defects, Les Houches, Session 35}. North--Holland, Amsterdam, 1981.

\bibitem{Marder59}
L.~Marder.
\newblock Flat space--times with gravitational fields.
\newblock {\em Proc. Roy. Soc. London, Ser. A}, 252:45--50, 1959.

\bibitem{Marder62}
L.~Marder.
\newblock Locally isometric spacetimes.
\newblock In {\em Recent Developments in General Relativity -- Volume Dedicated
  to {L}eopold {I}nfeld in Connection with his $60^{\rm th}$ Birthday}, pages
  333--338. Pergamon Press, Oxford, 1962.

\bibitem{Geroch87}
R.~Geroch and J.~Traschen.
\newblock Strings and other distributional sources in general relativity.
\newblock {\em Phys. Rev. D}, 36:1017--1031, 1987.

\bibitem{Clarke96}
C.J.S. Clarke, J.A. Vickers, and J.P. Wilson.
\newblock Generalized functions and distributional curvature of cosmic strings.
\newblock {\em Class. Quantum Grav.}, 13:2485--2498, 1996.

\bibitem{Gott85}
J.R. {Gott, III}.
\newblock Gravitational lensing effects of vacuum strings: exact solutions.
\newblock {\em Astrophys. J.}, 288:422--427, 1985.

\bibitem{Linet85}
B.~Linet.
\newblock The static metrics with cylindrical symmetry describing a model of
  cosmic strings.
\newblock {\em Gen. Rel. Grav}, 17:1109--1115, 1985.

\bibitem{Soleng90}
H.H. Soleng.
\newblock Spin-polarized cylinder in {E}instein--{C}artan theory.
\newblock {\em Class. Quantum Grav.}, 7:999--1007, 1990.

\bibitem{Soleng92a}
H.H. Soleng.
\newblock A spinning string.
\newblock {\em Gen. Rel. Grav.}, 24:111--117, 1992.

\bibitem{Sakharov67}
A.D. Sakharov.
\newblock Vacuum quantum fluctuations in curved space and the theory of
  gravitation.
\newblock {\em Doklady Akad.\ Nauk S.S.S.R.}, 177:70--71, 1967.
\newblock [{\em Sov. Physics Doklady}, 12:1040--1041, 1968].

\bibitem{Kroner90}
E.~Kr{\"o}ner.
\newblock The differential geometry of elementary point and line defects in
  {B}ravais crystals.
\newblock {\em Int. J. Theor. Phys.}, 29:1219--1237, 1990.

\bibitem{Mistura90}
L.~Mistura.
\newblock {C}artan connection and defects in {B}ravais lattices.
\newblock {\em Int. J. Theor. Phys}, 29:1207--1218, 1990.

\bibitem{Kadic83}
A.~Kadi{\'c} and D.G.B. Edelen.
\newblock {\em A Gauge Theory of Dislocations and Disclinations}, volume 174 of
  {\em Lecture Notes in Physics}.
\newblock Springer, Berlin, 1983.

\bibitem{Kroner86}
E.~Kr{\"o}ner.
\newblock On gauge theory in defect mechanics.
\newblock In E.~Kr{\"o}ner and K.~Kirchg{\"a}ssner, editors, {\em Trends in
  Applications of Pure Mathematics to Mechanics}, volume 249 of {\em Lecture
  Notes in Physics}, pages 281--294. Springer, Berlin, 1986.

\bibitem{Kleinert89}
H.~Kleinert.
\newblock {\em Gauge Fields in Condensed Matter}, volume~II.
\newblock World Scientific, Singapore, 1989.

\bibitem{Rivier90}
N.~Rivier.
\newblock Gauge theory and geometry of condensed matter.
\newblock In J.F. Sadoc, editor, {\em Geometry in Condensed Matter Physics},
  volume~9 of {\em Directions in Condensed Matter Physics}, pages 1--88. World
  Scientific, Singapore, 1990.

\bibitem{Malyshev93}
C.~Malyshev.
\newblock Underlying algebraic and gauge structures of the theory of
  disclinations.
\newblock {\em Arch. Mech.}, 45:93--105, 1993.

\bibitem{Tseytlin82}
A.A. Tseytlin.
\newblock {P}oincar\'e and de {S}itter gauge theories of gravity with
  propagating torsion.
\newblock {\em Phys. Rev. D}, 26:3327--3341, 1982.

\bibitem{Ivanenko83}
D.~Ivanenko and G.~Sardanashvily.
\newblock The gauge treatment of gravity.
\newblock {\em Phys. Rep.}, 94:1--45, 1983.

\bibitem{Mielke93a}
E.W. Mielke, J.D. McCrea, Y.~Ne'eman, and F.W. Hehl.
\newblock Avoiding degenerate coframes in an affine gauge approach to quantum
  gravity.
\newblock {\em Phys. Rev. D}, 48:673--679, 1993.

\bibitem{Hehl95}
F.W. Hehl, J.D. McCrea, E.W. Mielke, and Y.~Ne'eman.
\newblock Metric-affine gauge theory of gravity: Field equations, {N}oether
  identities, world spinors, and breaking of dilation invariance.
\newblock {\em Phys. Rep.}, 258:1--171, 1995.

\bibitem{Mielke87}
E.W. Mielke.
\newblock {\em Geometrodynamics of Gauge Fields}, volume~3 of {\em Physical
  Research}.
\newblock Akademie-Verlag, Berlin, 1987.

\bibitem{Grignani92}
G.~Grignani and G.~Nardelli.
\newblock Gravity and the {P}oincar\'e group.
\newblock {\em Phys. Rev. D}, 45:2719--2731, 1992.

\bibitem{Trautman79}
A.~Trautman.
\newblock The geometry of gauge fields.
\newblock {\em Czech. J. Phys. B}, 29:107--116, 1979.

\bibitem{Yang74}
C.N. Yang.
\newblock Integral formalism for gauge fields.
\newblock {\em Phys. Rev. Lett.}, 33:445--447, 1974.

\bibitem{Bezerra91}
V.B. Bezerra and P.S. Letelier.
\newblock Holonomy transformations and a class of spacetime topological
  defects.
\newblock {\em Class. Quantum Gravity}, 8:L73--L76, 1991.

\bibitem{Petti86}
R.J. Petti.
\newblock On the local geometry of rotating matter.
\newblock {\em Gen. Rel. Grav.}, 18:441--461, 1986.

\bibitem{Gockeler87}
M.~G{\"o}ckeler and T.~Sch{\"u}cker.
\newblock {\em Differential Geometry, Gauge Theories, and Gravity}.
\newblock Cambridge Monographs in Mathematical Physics. Cambridge University
  Press, Cambridge, 1987.

\bibitem{Trautman73a}
A.~Trautman.
\newblock On the structure of the {E}instein--{C}artan equations.
\newblock In {\em Differential Geometry, Symposia Mathematica}, volume~12,
  pages 139--162. Academic Press, London, 1973.

\bibitem{Baekler92}
P.~Baekler, E.W. Mielke, and F.W. Hehl.
\newblock Dynamical symmetries in topological {3D} gravity with torsion.
\newblock {\em Nuovo Cimento B}, 107:91--110, 1992.

\bibitem{Hehl74}
F.W. Hehl, P.~Von der Heyde, and G.D. Kerlick.
\newblock General relativity with spin and torsion and its deviations from
  {E}instein's theory.
\newblock {\em Phys. Rev. D}, 10:1066--1069, 1974.

\bibitem{Kerlick75}
G.D. Kerlick.
\newblock Cosmology and particle production via gravitaitonal spin--spin
  interaction in the {E}instein--{C}artan--{S}ciama--{K}ibble theory of
  gravity.
\newblock {\em Phys. Rev. D}, 12:3004--3006, 1975.

\bibitem{Smalley94}
L.S. Smalley and J.P. Krisch.
\newblock Energy conditions in non-{R}iemannian spacetimes.
\newblock {\em Phys. Lett. A}, 196:147--153, 1994.

\bibitem{Obukhov93}
Yu.N. Obukhov and R.~Tresguerres.
\newblock Hyperfluid --- a model of classical matter with hypermomentum.
\newblock {\em Phys. Lett. A}, 184:17--22, 1993.

\bibitem{Obukhov96}
Yu.N. Obukhov.
\newblock On a model of an unconstrained hyperfluid.
\newblock {\em Phys. Lett. A}, 210:163--167, 1996.

\bibitem{Arkuszewski75}
W.~Arkuszewski, W.~Kopczy{\'n}ski, and V.N. Ponomariev.
\newblock Matching conditions in the {E}instein--{C}artan theory of
  gravitation.
\newblock {\em Commun. Math. Phys.}, 45:183--190, 1975.

\bibitem{Chmielowski92}
P.~Chmielowski.
\newblock Matching and junction conditions in the {P}oincar\'e gauge theory of
  gravity with quadratic Lagrangian.
\newblock {\em Prog. Theor. Phys.}, 88:691--702, 1992.

\bibitem{Soleng:PRD94}
H.H.~Soleng.
\newblock Negative energy densities in extended sources generating closed timelike curves in 
general relativity with and without torsion. 
\newblock {\em Phys. Rev. D}, 49:1124--1125, 1994.

\bibitem{Mazur86}
P.O. Mazur.
\newblock Induced angular momentum on superconducting cosmic strings.
\newblock {\em Phys. Rev. D}, 34:1925--1927, 1986.

\bibitem{Ray82}
J.R. Ray and L.R. Smalley.
\newblock Improved perfect-fluid energy--momentum tensor with spin in
  {E}instein--{C}artan spacetime.
\newblock {\em Phys. Rev. Lett.}, 49:1059--1061, 1982.
\newblock Erratum: {\it Phys. Rev. Lett.}, 50:623, 1983.

\bibitem{Krisch96}
J.P. Krisch.
\newblock A family of strings with spin.
\newblock {\em Gen. Rel. Grav.}, 28:69--76, 1996.

\bibitem{Hearn93}
A.C. Hearn.
\newblock {\em {REDUCE} User's Manual, Version 3.5}.
\newblock RAND Publication CP78 (Rev. 10/93), The RAND Corporation, Santa
  Monica, CA 90407--2138, 1993.

\bibitem{Schrufer94}
E.~Schr\"ufer.
\newblock {\em EXCALC: A System for Doing Calculations in the Calculus of
  Modern Differential Geometry}.
\newblock GMD, Institut I1, D--53757 St.Augustin, Germany, 1994.

\bibitem{Soleng94}
H.H. Soleng.
\newblock {\em Tensors in Physics: User's Guide and Disk for the 
  Mathematica Package {\sc Cartan}, Version 1.2}.
\newblock Scandinavian University Press, Oslo, 1996.

\end{thebibliography}
\end{document}